\documentclass{article}
\pdfoutput=1
\usepackage{PRIMEarxiv}
\usepackage[utf8]{inputenc} 
\usepackage[T1]{fontenc}    
\usepackage{hyperref}       
\usepackage{url}            
\usepackage{booktabs}       
\usepackage{nicefrac}       
\usepackage{microtype}      
\usepackage{lipsum}
\usepackage{fancyhdr}       

\usepackage[pdftex]{graphicx} 
\usepackage{amsmath,amsfonts,amssymb,mathtools}
\usepackage{mathrsfs}
\newcommand{\upi}{\mathrm{i}}
\newcommand{\upd}{\mathrm{d}}
\newcommand{\intinf}{\int_{-\infty}^\infty}

\newcommand{\Rtwo}{{\mathbb{R}^2}}
\DeclareMathAlphabet{\mathsfbi}{OT1}{\sfdefault}{bx}{sl}
\renewcommand{\Re}{\mathrm{Re}}
\renewcommand{\Im}{\mathrm{Im}}
\newcommand{\bx}{\mathbf{x}}
\newcommand{\Z}{\mathbb{Z}}
\usepackage{enumitem}

\title{Generalised eigenfunction expansion and singularity expansion methods for two-dimensional acoustic time-domain wave scattering problems}

\author{
  Ben Wilks\\
  Department of Mathematics and Statistics\\
  University of Otago\\
  Dunedin\\
  New Zealand\\
  \texttt{wilbe612@student.otago.ac.nz}\\
  \AND
  Michael H. Meylan \\
  School of Information and Physical Sciences\\
  The University of Newcastle\\
  Newcastle\\
  Australia\\
  \AND
  Fabien Montiel\\
  Department of Mathematics and Statistics\\
  University of Otago\\
  Dunedin\\
  New Zealand\\
  \AND
  Sarah Wakes\\
  Department of Mathematics and Statistics\\
  University of Otago\\
  Dunedin\\
  New Zealand\\
}

\begin{document}
\maketitle

\begin{abstract}
Time-domain wave scattering in an unbounded two-dimensional acoustic medium by sound-hard scatterers is considered. Two canonical geometries, namely a split-ring resonator (SRR) and an array of cylinders, are used to highlight the theory, which generalises to arbitrary scatterer geometries. The problem is solved using the generalised eigenfunction expansion method (GEM), which expresses the time-domain solution in terms of the frequency-domain solutions. A discrete GEM is proposed to numerically approximate the time-domain solution. It relies on quadrature approximations of continuous integrals and can be thought of as a generalisation of the discrete Fourier transform. The solution then takes a simple form in terms of direct matrix multiplications.
In parallel to the GEM, the singularity expansion method (SEM) is also presented and applied to the two aforementioned geometries. It expands the time-domain solution over a discrete set of unforced, complex resonant modes of the scatterer. Although the coefficients of this expansion are divergent integrals, we introduce a method of regularising them using analytic continuation. The results show that while the SEM is usually inaccurate at $t=0$, it converges rapidly to the GEM solution at all spatial points in the computational domain, with the most rapid convergence occurring inside the resonant cavity.
\end{abstract}

\section{Introduction}
The majority of contemporary linear wave scattering research focusses exclusively on the frequency-domain problem (i.e.\ the problem of finding time-harmonic solutions of the associated wave equation). These problems and the numerical techniques for solving them are very well understood \cite{martin2006}. The lack of research into the corresponding time-domain scattering problems is unfortunate because features such as resonances and decay patterns, which can characterise a transient response, can be difficult to discern from the frequency-domain solutions alone. Recently, we described two methods to solve the time-domain wave scattering problem by applying them to the prototypical one-dimensional wave scattering problem by a point scatterer on an infinitely stretched string \cite{wilks2023canonical}. This paper seeks to extend that work to two-dimensional wave-scattering problems.

The first method we consider is the generalised eigenfunction expansion method (GEM). This method, which was introduced by Povzner \cite{povzner1953expansion} and further developed by Ikebe \cite{ikebe1960eigenfunction} for the Schr\"{o}dinger equation, and later by Hazard and Loret in the context of hydrodynamics \cite{hazard2007generalized}, uses spectral theory to expand the time-domain solution in terms of the frequency-domain solutions. In particular, the frequency-domain solutions are eigenfunctions of the operator which describes the temporal evolution. Therefore, the GEM directly maps the frequency domain to the time domain. The connection between the time and frequency domains is less clear in more well-known methods, such as those based on the Laplace transform \cite{martin2021time}. The GEM can be applied numerically by approximating the integrals appearing in its derivation using quadrature \cite{meylan_2009,meylan2009time}. We term the resulting numerical method the \textit{discrete GEM}, because it is related to the GEM in the same way that the discrete Fourier transform is related to the Fourier transform. An important insight that was described in \cite{Meylan2023MWSW03} is that the discrete GEM solution to the time-domain problem can be implemented efficiently in terms of matrix multiplication. The extension to the GEM to wave scattering problems in two-dimensions, which brings additional technical difficulties to those described in \cite{wilks2023canonical}, is described here.

Following our previous work \cite{wilks2023canonical}, we study the singularity expansion method (SEM) alongside the GEM. The SEM, which was developed by Baum \cite{baum1971singularity,baum2005singularity}, approximates the time-domain solution as an expansion over a resonator's unforced modes, which we refer to as complex resonances. These are nontrivial solutions to the scattering problem that occur in the absence of forcing at a single complex frequency \cite{pagneux2013trapped,meylan2017extraordinary}. With the exception of trapped modes (which are special cases of complex resonances that occur at real frequencies and have finite energy \cite{LINTON200716}), complex resonances are unbounded in the far field. These must be normalised in order to obtain the coefficients of the SEM expansion, although doing so leads to divergent integrals due to the resonant modes being unbounded. Fortunately, it is known that these modes can be normalised by regularising these divergent integrals (i.e.\ using analytic continuation) \cite{Kristensen2020}. In this paper, we show how to regularise these integrals in a two-dimensional wave-scattering context, provided that the complex resonances can be expanded in terms of Hankel functions sufficiently far from the scatterer. We are not aware of any similar application of the SEM to a two-dimensional scattering problem in free space; thus we believe our regularisation approach is novel.

The context of this paper is acoustic wave scattering in two dimensions. However, since the underlying equation is the two-dimensional wave equation, the methods and results apply more generally (e.g.\ to shallow water hydrodynamics \cite{stoker1957water}). As examples, two resonant scatterers are considered, namely a split-ring resonator (SRR) and an array of sound-hard cylinders. The resonant properties of both of these have been studied extensively in the frequency domain. SRRs, which consist of a cylindrical shell with one or more openings, are often proposed as the subunits of metamaterials in a wide range of applications, including for electromagnetic waves, acoustics and water waves \cite{Smith2000,llewellyn2010,Hu2011,krynkin2011scattering,montiel2017,Bennetts2018,montiel2020,smith2022}. The resonances of arrays of sound-hard circular cylinders (and the related problem of arrays of bottom-mounted circular cylinders immersed in water) have also been considered in multiple configurations, including square arrays \cite{evans1999trapping,meylan2009time}, line arrays \cite{maniar1997wave,eatock2007modelling,thompson2008new,bennetts2022rayleigh} and circular arrays \cite{EVANS199783,maling2016whispering}. While this paper only considers these two examples, the methods we present generalise to scatterers of arbitrary geometry.

The outline of this paper is as follows. The acoustic scattering problem is outlined in \textsection\ref{Preliminaries_sec}, after which the GEM is introduced in \textsection\ref{GEM_sec}. In \textsection\ref{frequency_domain_sec}, we solve the frequency-domain problems associated with the SRR and the array of cylinders. The discrete GEM is then described in \textsection\ref{discrete_GEM_sec}, in which we show that it can be expressed in terms of matrix multiplication. In \textsection\ref{SEM_sec}, we introduce the SEM and explain how to regularise the divergent normalisation integrals. The results are presented in \textsection\ref{results_sec}, which show excellent qualitative agreement between the GEM and SEM. Brief concluding remarks are given in \textsection\ref{conclusion_sec}.

\section{Preliminaries}\label{Preliminaries_sec}
We consider the scattering of acoustic waves by a bounded sound-hard scatterer $\Gamma$ in an acoustic medium in free space. This framework can be extended to allow for scatterers with varying material properties or to sound soft scatterers, but this is not considered here. We assume that all quantities are homogeneous in one spatial direction (i.e., the $z$-direction) so that the underlying partial differential equation can be written in two spatial dimensions (with corresponding Cartesian coordinates $x$ and $y$). Assuming low amplitude fluctuations in pressure and density, the initial boundary-value problem is described by
\begin{subequations} \label{wave_eq}
\begin{align}
    \frac{\partial^2\phi}{\partial t^2}&=c^2\bigtriangleup\phi&\bx\in\Omega \label{wave_eq1}\\
    \partial_n\phi&=0&\bx\in\partial\Gamma\\
    \phi(\bx,0)&=f(\bx)&\bx\in\Omega\\
    \partial_t\phi(\bx,0)&=g(\bx)&\bx\in\Omega.
\end{align}
\end{subequations}
Here, $\phi$ is the velocity potential, which is related to the relative pressure $p$ and the particle velocity $\mathbf{v}$ via
\begin{equation}
    p=-\rho\partial_t\Phi\quad\text{and}\quad\mathbf{v}=\nabla\Phi,
\end{equation}
respectively, where $\rho$ is the equilibrium density of the acoustic medium \cite{rossing2015springer}. The speed of sound is given by $c=\sqrt{B_0/\rho}$, where $B_0$ is the bulk modulus of the acoustic medium, which occupies the region $\Omega\coloneqq\Rtwo\setminus\Gamma$. We have also introduced $\bx=(x,y)$ and $\partial\Gamma$ denotes the boundary of $\Gamma$.

We seek a solution to \eqref{wave_eq} by rewriting \eqref{wave_eq1} as a first-order linear system
\begin{equation}\label{two_component_TD}
    \upi\partial_t\begin{bmatrix}
        \phi\\\upi\eta
    \end{bmatrix}
    =\mathscr{P}\begin{bmatrix}
        \phi\\\upi\eta
    \end{bmatrix},
\end{equation}
where
\begin{equation}
    \mathscr{P}=\begin{bmatrix}
        0&1\\-c^2\bigtriangleup&0
    \end{bmatrix}.
\end{equation}
We introduce the following energy inner product
\begin{equation}\label{energy_inner_product}
    \left\langle\begin{bmatrix}
        \phi_1\\\upi\eta_1
    \end{bmatrix},\begin{bmatrix}
        \phi_2\\\upi\eta_2
    \end{bmatrix}\right\rangle=\int_{\Omega}\frac{1}{c^2}\eta_1\eta_2^*+\nabla\phi_1\cdot\nabla\phi_2^*\upd\bx.
\end{equation}
which is related to the total energy of the acoustic wave (being the sum of the kinetic and potential energy \cite{rossing2015springer}) by
\begin{equation}
    E=\frac{\rho_0}{2}\left\langle\begin{bmatrix}
        \phi\\\upi \eta
    \end{bmatrix}(\cdot,t),\begin{bmatrix}
        \phi\\\upi \eta
    \end{bmatrix}(\cdot,t)\right\rangle=\frac{\rho_0}{2}\left\langle\begin{bmatrix}
        f\\\upi g
    \end{bmatrix},\begin{bmatrix}
        f\\\upi g
    \end{bmatrix}\right\rangle,
\end{equation}
where the second equality follows from conservation of energy (in other words, energy at time $t$ is equal to energy at time zero). The function space associated with the inner product \eqref{energy_inner_product} is
\begin{equation}\label{function_space}
    \left\{\left.\begin{bmatrix}
        \phi\\\upi\eta
    \end{bmatrix}:\Omega\to\mathbb{C}^2\right|\left\langle\begin{bmatrix}
        \phi\\\upi\eta
    \end{bmatrix},\begin{bmatrix}
        \phi\\\upi\eta
    \end{bmatrix}\right\rangle<\infty\quad\text{and}\quad \partial_n\phi=0\text{ on }\partial\Gamma\right\}.
\end{equation}

\section{The generalised eigenfunction expansion method}\label{GEM_sec}
The operator $\mathscr{P}$ is self-adjoint with respect to the inner product \eqref{energy_inner_product}. To show this, we compute
\begin{align}
    \left\langle\mathscr{P}\begin{bmatrix}
        \phi_1\\\upi\eta_1
    \end{bmatrix},\begin{bmatrix}
        \phi_2\\\upi\eta_2
    \end{bmatrix}\right\rangle&=\int_\Omega \upi(\bigtriangleup\phi_1)(\eta_2^*)+(\nabla\upi\eta_1)\cdot(\nabla\phi_2^*)\upd\bx\nonumber\\
    &=\int_\Omega\nabla\phi_1\cdot\nabla(\upi\eta_2)^*+\frac{1}{c^2}\eta_1(\upi c^2\bigtriangleup\phi_2)^*\upd \bx\nonumber\\
    &=\left\langle\begin{bmatrix}
        \phi_1\\\upi\eta_1
    \end{bmatrix},\mathscr{P}\begin{bmatrix}
        \phi_2\\\upi\eta_2
    \end{bmatrix}\right\rangle,
\end{align}
where the second line follows from two applications of Green's first identity. As a consequence of being self-adjoint, $\mathscr{P}$ has real eigenvalues $\omega$, that is,
\begin{equation}\label{eigenfunction_equation}
    \begin{bmatrix}
        0&1\\-c^2\bigtriangleup&0
    \end{bmatrix}\begin{bmatrix}
        \breve{\phi}_m\\\upi\breve{\eta}_m
    \end{bmatrix}(\cdot,\omega)=\omega\begin{bmatrix}
        \breve{\phi}_m\\\upi\breve{\eta}_m
    \end{bmatrix}(\cdot,\omega),
\end{equation}
where $\{[\breve{\phi}_m,\upi\breve{\eta}_m]^\intercal(\cdot,\omega)|m\in\mathbb{Z}\}$ is assumed to be an orthogonal basis of the eigenspace of $\mathscr{P}$ at the fixed frequency $\omega$ (note that eigenfunctions at two different eigenvalues $\omega$ and $\omega^\prime$ are automatically orthogonal because $\mathscr{P}$ is self-adjoint). Following \cite{meylan2009time}, we choose such a basis by taking $[\breve{\phi}_m,\upi\breve{\eta}_m]^\intercal$ to be the solution of the scattering problem in which $J_m(kr)e^{\upi m\theta}$, $m\in\mathbb{Z}$, describes the incident wave, where $(r,\theta)$ is the system of polar coordinates for the problem, $k=|\omega|/c$ and $J_m$ is the Bessel function of the first kind of order $m$. This scattering problem can be formulated as
\begin{subequations}\label{freq_dom_eqs}
\begin{align}
\bigtriangleup\breve{\phi}_m(\bx,\omega)&=-k^2\breve{\phi}_m(\bx,\omega)&\bx\in\Omega\label{helmholtz}\\
\partial_n\breve{\phi}_m(\bx,\omega)&=0&\bx\in\partial\Gamma\label{Neumann_BC_FD}\\
\sqrt{r}(\partial_r-\upi k)(\breve{\phi}_m(\bx,\omega)-J_m(kr)e^{\upi m\theta}) &\to 0&\text{ as }r\to\infty\label{Sommerfeld}\\
\breve{\eta}_m(\bx,\omega)&=-\upi\omega\breve{\phi}_m(\bx,\omega),
\end{align}
\end{subequations}
in which the Helmholtz equation \eqref{helmholtz} results from combining the two components of \eqref{eigenfunction_equation} into a single equation, and the requirement that the scattered wave $\breve{\phi}_m-J_m(kr)e^{\upi m\theta}$ is outgoing is enforced by \eqref{Sommerfeld}. Note that $[\breve{\phi}_m,\upi\breve{\eta}_m]^\intercal$ are termed \textit{generalised} eigenfunctions because they have infinite energy, i.e.\ they are not in the function space described by \eqref{function_space}.

In the absence of any trapped modes, the continuous spectrum of $\mathscr{P}$ is the entire real line. The spectral theorem then allows us to expand the time-domain solution as a continuous superposition of the frequency-domain solutions
\begin{equation}\label{FD_expansion0}
    \begin{bmatrix}
        \phi\\\upi \eta
    \end{bmatrix}(\bx,t)=\intinf\sum_{m=-\infty}^\infty \widetilde{A}_m(\omega,t)\begin{bmatrix}
        \breve{\phi}_m\\\upi\breve{\eta}_m
    \end{bmatrix}(\bx,\omega)\upd\omega.
    \end{equation}
By substituting \eqref{FD_expansion0} into \eqref{two_component_TD} and applying the orthogonality of the frequency-domain solutions, we find that time dependence of the coefficients of \eqref{FD_expansion0} can be factored as $\widetilde{A}_m(\omega,t)=A_m(\omega)e^{-\upi\omega t}$, thus
\begin{equation}\label{FD_expansion}
    \begin{bmatrix}
        \phi\\\upi \eta
    \end{bmatrix}(\bx,t)=\intinf\sum_{m=-\infty}^\infty A_m(\omega)\begin{bmatrix}
        \breve{\phi}_m\\\upi\breve{\eta}_m
    \end{bmatrix}(\bx,\omega)e^{-\upi\omega t}\upd\omega,
    \end{equation}
Since the frequency-domain solutions have time-reversal symmetry (by which $\breve{\phi}_m(\bx,-\omega)=\breve{\phi}_m(\bx,\omega)^*$) and since $\phi(\bx,t)$ is real, the spectral amplitudes must satisfy $A_m(-\omega)=A_m(\omega)^*$ and \eqref{FD_expansion} reduces to
    \begin{equation}\label{FD_expansion2}
    \begin{bmatrix}
        \phi\\\upi \eta
    \end{bmatrix}(\bx,t)=2\,\Re\left\{\int_0^\infty\sum_{m=-\infty}^\infty A_m(\omega)\begin{bmatrix}
        \breve{\phi}_m\\\upi\breve{\eta}_m
    \end{bmatrix}(\bx,\omega)e^{-\upi\omega t}\upd\omega\right\}.
    \end{equation}    
Next, we compute the normalisation of the frequency-domain solutions, i.e.\ the values $\widetilde{N}_m(\omega)$, where
\begin{equation}
    \left\langle\begin{bmatrix}
        \breve{\phi}_m\\\upi\breve{\eta}_m
    \end{bmatrix}(\cdot,\omega),\begin{bmatrix}
        \breve{\phi}_l\\\upi\breve{\eta}_l
    \end{bmatrix}(\cdot,\omega^\prime)\right\rangle=\widetilde{N}_m(\omega)\delta(\omega-\omega^\prime)\delta_{ml}.
\end{equation}
To do this, we assume that the normalisation is identical to the case where no scatterers are present---a fact that generally appears to hold \cite{povzner1953expansion,ikebe1960eigenfunction,wilcox1975scattering,hazard2002,hazard2007generalized,wilks2023canonical}. This allows us to use the Hankel transform to compute $\widetilde{N}_m(\omega)=4\pi\omega$. The spectral amplitudes in \eqref{FD_expansion2} are computed from the initial conditions
\begin{align}
A_m(\omega)&=\frac{1}{4\pi\omega}
    \left\langle\begin{bmatrix}
        f\\\upi g
    \end{bmatrix},\begin{bmatrix}
        \breve{\phi}_m\\\upi \breve{\eta}_m
    \end{bmatrix}(\cdot,\omega)\right\rangle\nonumber\\
    &=\frac{1}{4\pi\omega}
   \int_{\Omega}\frac{1}{c^2}g(\bx)\breve{\eta}_m^*(\bx,\omega)+k^2f(\bx)\breve{\phi}_m^*(\bx,\omega)\upd\bx\nonumber\\
   &=\frac{1}{4\pi c^2}
   \int_{\Omega}\left(\upi g(\bx)+\omega f(\bx)\right)\breve{\phi}_m^*(\bx,\omega)\upd\bx,
   \label{GEM_spectrum}
\end{align}
where the first equality follows from applying orthogonality to \eqref{FD_expansion}. The second equality is a consequence of Green's first identity and the fact that $\breve{\phi}_m(\cdot,\omega)$ satisfies the Helmholtz equation \eqref{helmholtz}. The third equality follows from \eqref{eigenfunction_equation}.

Equations \eqref{FD_expansion2} and \eqref{GEM_spectrum} describe the general form of the time-domain solution in terms of the frequency-domain solutions, which we must now calculate. In the next section, we give the frequency-domain solutions for two canonical problems. However, equations \eqref{FD_expansion2} and \eqref{GEM_spectrum} apply to any two-dimensional linear wave scattering problem by a bounded, non-dissipative scatterer in a homogeneous background medium, provided that the frequency-domain solutions can be computed.

\section{Frequency-domain problems}\label{frequency_domain_sec}
In this section, we derive solutions to \eqref{freq_dom_eqs} for two canonical scatterers $\Gamma$, namely the SRR and an array of sound hard cylinders. The former problem is solved using an integral equation/Galerkin method, while the latter is solved using self-consistent multiple scattering theory. In both cases, we truncate the infinite series solutions to obtain a matrix equation of the form
\begin{equation}\label{classic_matrix_eq}
    \mathcal{M}(\omega)\mathbf{\widetilde{x}}=\mathbf{\widetilde{f}},
\end{equation}
where $\mathcal{M}(\omega)$ is the frequency-dependent matrix, $\mathbf{\widetilde{f}}$ is obtained from the incident potential and $\mathbf{\widetilde{x}}$ is a vector of coefficients associated with the scattered potential. Matrix equations of this form appear universally in frequency domain wave scattering problems, for a wide range of numerical methods, e.g.\ Galerkin methods \cite{Porter1995a,montiel2017}, eigenfunction expansion methods \cite{meylan2017extraordinary}, self-consistent multiple scattering methods \cite{martin2006,peter_meylan_linton_2006,montiel2015Evolution}, the boundary element method \cite{Amini1995,WANG2004557}, the finite element method \cite{masmoudi1987,kirsch1990} or others.

\subsection{The split ring resonator}
\label{split_ring_solution}
\begin{figure}
    \centering
    \includegraphics[width=0.3\textwidth]{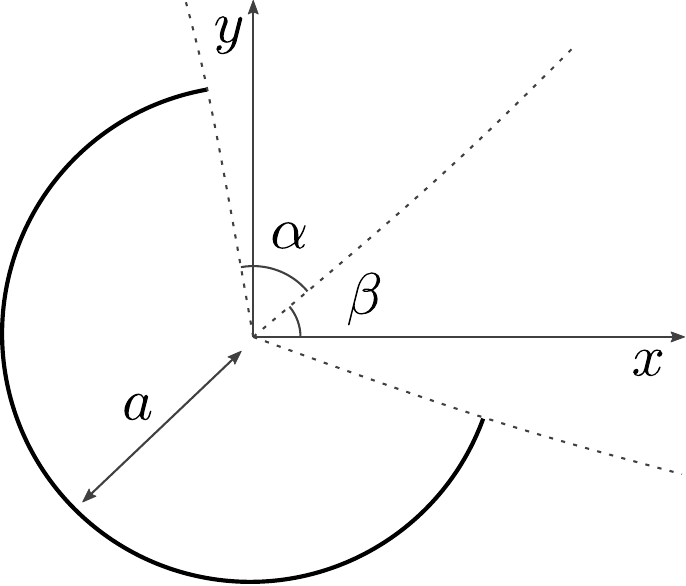}
    \caption{Schematic of the SRR. The radius of the SRR is denoted $a$, its opening angle is given by $2\alpha$, and its orientation with respect to the positive $x$-axis is denoted $\beta$.}
    \label{fig:split-ring}
\end{figure}

We now consider the scattering by a SRR, which is described by
\begin{equation}
    \Gamma = \left\{\left.(a\cos s,a\sin s)\right|s\in(\alpha+\beta,2\pi-\alpha+\beta)\right\}.
\end{equation}
Here, $\alpha\in(0,\pi)$ is half of the opening angle of the SRR in radians and $a>0$ is the radius in metres. The orientation of the centre of the opening from the positive $x$ axis in radians is denoted by $\beta$. A schematic of the SRR is given in figure \ref{fig:split-ring}. To solve this problem, we use separation of variables to obtain multipole expansions for the velocity potential $\breve{\phi}_m$ in polar coordinates, then compute the unknown coefficients using an integral equation/Galerkin method. Most of the derivations are identical to those of Montiel et al. \cite{montiel2017}, where the reader is directed for further details, although we have made several modifications.

The unknown potential $\breve{\phi}_m$ is separated into its exterior and interior subdomains as
\begin{equation}
    \breve{\phi}_m(x,\omega) = \begin{cases} \breve{\phi}_{\mathrm{Ext}}&r>a\\
    \breve{\phi}_{\mathrm{Int}}&r<a.
    \end{cases}
\end{equation}
After accounting for the boundary conditions, the Sommerfeld radiation condition, the incident wave $J_m(kr)e^{\upi n \theta}$ and the requirement that $\breve{\phi}_m$ is nowhere singular, separation of variables allows us to write
\begin{subequations}\label{SRR_expansions}
\begin{align}
\breve{\phi}_{\mathrm{Ext}}&=J_m(kr)e^{\upi m\theta}+\phi_{\mathrm{Sc}}\\
\breve{\phi}_{\mathrm{Sc}}(r,\theta) &= \sum_{n=-\infty}^\infty C_n\frac{H_n^{(1)}(kr)}{H_n^{(1)\prime}(ka)}e^{\upi n\theta},\label{SRR_scattered}\\
\breve{\phi}_{\mathrm{Int}}(r,\theta) &= \sum_{n=-\infty}^\infty D_n\frac{J_n(kr)}{J_n^\prime(ka)}e^{\upi n\theta},\label{SRR_interior}
\end{align}
\end{subequations}
where the coefficients $C_n$ and $D_n$ are unknown. The potential of the scattered wave is denoted $\breve\phi_{\mathrm{Sc}}$. We have defined $J_n$ and $H_n^{(1)}$ as Bessel and Hankel functions of the first kind of order $n$, whose derivatives with respect to their argument are denoted by $J_n^\prime$ and $H_n^{(1)\prime}$, respectively.

By imposing the Neumann boundary condition \eqref{Neumann_BC_FD} on $\Gamma$ and requiring that $\breve{\phi}_m$ is continuously differentiable across the gap $\{(a,\theta):\beta-\alpha<\theta<\beta+\alpha\}$, the scattering problem is converted into an integral equation for an unknown auxiliary function $u(\theta)$, which is defined as
\begin{equation}
    u(\theta)=\partial_r \breve{\phi}_m(a,\theta).
\end{equation}
The integral equation is solved using a Galerkin method, which aims to determine the coefficients $B_p$ that satisfy the approximation. 
\begin{equation}
    u(\theta)\approx\sum_{p=1}^{N_{\mathrm{aux}}} B_p v_p(\theta)
\end{equation}
for $\theta\in(\beta-\alpha,\beta+\alpha)$, where the auxiliary basis functions $v_p$ will be defined later. We obtain the coefficients $B_p$ by solving a matrix equation of the form
\begin{equation}\label{linear_system_for_split_ring}
    \mathcal{M}(\omega)\mathbf{B}=2\pi k UK\,\mathrm{Diag}(e^{\upi n\beta})\mathbf{F},
\end{equation}
where $\mathcal{M}(\omega)=\mathcal{UKU}^\dag$, in which $\dag$ denotes the conjugate transpose. The vector $\mathbf{F}$ is of length $2N_{\mathrm{sol}}+1$ and contains the terms $F_n\coloneqq\delta_{mn}J_m^\prime(ka)$ for $-N_{\mathrm{sol}}<n<N_{\mathrm{sol}}$, where $\delta_{mn}$ is the Kronecker delta. The vector $\mathbf{B}$ is of length $N_{\mathrm{aux}}$ and contains the coefficients $B_p$. The square diagonal matrices $\mathcal{K}$ and $K$ are of size $(2N_{\mathrm{ker}}+1)\times(2N_{\mathrm{ker}}+1)$ and $(2N_{\mathrm{sol}}+1)\times(2N_{\mathrm{sol}}+1)$ respectively, each with diagonal entries
\begin{equation}\label{wronskian}
    \mathcal{K}_{n} = \frac{2\upi}{\pi k a J_n^\prime(ka)H_n^{(1)\prime}(ka)}.
\end{equation}
When $n$ is large, the numerical evaluation of $\mathcal{K}_{n}$ can fail, although this can be overcome by approximating the Bessel function and Hankel function derivatives using their asymptotic forms for large order \cite{montiel2020}. The diagonal matrix $\mathrm{Diag}(e^{\upi n\beta})$ appearing in \eqref{linear_system_for_split_ring} encodes the rotation of the SRR, as was derived in the appendix of \cite{montiel2020}. Lastly, the matrices $\mathcal{U}$ and $U$ appearing in \eqref{linear_system_for_split_ring} are of size $N_{\mathrm{aux}}\times(2N_{\mathrm{ker}}+1)$ and $N_{\mathrm{aux}}\times(2N_{\mathrm{sol}}+1)$, respectively. The entries of these matrices are the integrals
\begin{equation}\label{inner_product_integrals}
    \mathcal{U}_{pn}=\int_{-\alpha}^\alpha v_p(\theta) e^{-\upi n \theta} \upd\theta.
\end{equation}
In order to anticipate the Meixner singularities of the order $-1/2$ occurring at the tips of the SRR (i.e.\ at $(r,\theta)=(a,\beta\pm\alpha)$), we choose the auxiliary basis functions to be
\begin{equation}\label{auxiliary_chebyshev}
    v_p(\theta) = \frac{1}{\sqrt{\alpha^2-\theta^2}}T_{p-1}(\theta/\alpha),
\end{equation}
where $T_l(x)$ is the Chebyshev polynomial of the first kind of order $l$. This basis was previously applied to a related problem of acoustic scattering by concentric SRRs by \cite{llewellyn2010}. Because this basis accounts for the Meixner singularities, we require fewer basis terms than the method of Montiel et al. \cite{montiel2017}, which used a Fourier basis, for the same degree of precision.

Having chosen the basis for the auxiliary function, the integrals in \eqref{inner_product_integrals} can be computed using equation 3.3.2 in \cite{bateman1954tables}, which yields
\begin{equation}
    \mathcal{U}_{pn}=\upi^{-(p-1)} \pi J_{p-1}(n\alpha).
\end{equation}
The unknown coefficients are recovered using the formulae
\begin{subequations}\label{recover_C_and_D}
    \begin{align}
    D_n&=\frac{1}{2\pi k}e^{-\upi n \beta}\sum_{p=-N_{\mathrm{aux}}}^{N_{\mathrm{aux}}}\mathcal{U}_{pn}^*B_p\\
    C_n&=D_n-F_n.\label{recover_C_D_2}
\end{align}
\end{subequations}

To generate the results in \textsection\ref{results_sec}, we choose $N_\mathrm{sol}=60$, $N_\mathrm{aux}=30$ and $N_\mathrm{ker}=2500$. These truncation parameter values were chosen to be sufficiently large so that at the points $\bx=(x,y)=(0,0)$ and $\bx=(-2,0)$, $\breve{\phi}_m(\bx,\omega)$ has converged to 4 decimal places for all $|m|<30$, when sampled at 80 frequency points spanning $\omega\in[0.1,20]$. In particular, the ratio between $N_\mathrm{sol}$ and $N_\mathrm{aux}$ has been optimised in preliminary tests. The accuracy of the method can be improved most readily by increasing $N_\mathrm{ker}$, which was also the case in similar numerical schemes \cite{Porter1995a}. 

\subsection{An array of circular cylinders}
\label{multiple_scattering_soln}
 We now consider the problem of planar acoustic scattering by $N$ sound-hard circular cylinders of radius $a$. This multiple scattering problem has been studied by many authors, but we recount the solution here for completeness. See \cite{martin2006} for further details.
 
\begin{figure}
    \centering
    \includegraphics[width=0.5\textwidth]{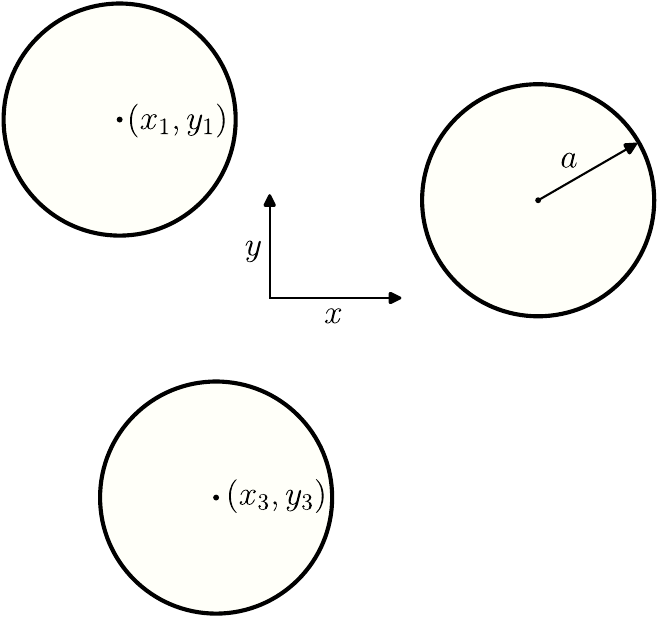}
    \caption{Schematic of an array of $N=3$ circular cylinders of radius $a$. The centre-points of the cylinders are denoted by $(x_j,y_j)$ for $j=1,\dots,N$.}
    \label{fig:cylinders_schematic}
\end{figure}

In two dimensions, we denote the centre-points of the cylinders as $(x_j,y_j)$, which we express in polar coordinates with respect to the origin as $(R_{0j},\varphi_{0j})$. The region occupied by the cylinders is $\Gamma=\cup_{j=1}^N\Gamma_j$, where
\begin{equation}
    \Gamma_j = \left\{(x,y)|(x-x_j)^2+(y-y_j)^2<a^2\right\}.
\end{equation}
A schematic of this problem for $N=3$ is given in figure \ref{fig:cylinders_schematic}

We introduce the local polar coordinates $(r_j,\theta_j)$ in order to solve the multiple scattering problem. These are defined by $x-x_j=r_j\cos\theta_j$ and $y-y_j=r_j\sin\theta_j$ for all $j=1,\dots,N$. We assume that the velocity potential can be described using a multipole expansion of the form
\begin{equation}\label{multipole1}
    \breve{\phi}_m(\bx,\omega) = J_m(kr)e^{\upi m \theta}+\sum_{j=1}^N \breve{\phi}_{\mathrm{Sc}}^{[j]},
\end{equation}
where $\phi_{\mathrm{Sc}}^{[j]}$ describes the wave scattered by the $j$th cylinder. These functions can be expressed in local polar coordinates as
\begin{align}
    \breve{\phi}_{\mathrm{Sc}}^{[j]}(r_j,\theta_j)=\sum_{n=-\infty}^\infty C_n^{[j]}\frac{H_n^{(1)}(kr_j)}{H_n^{(1)\prime}(ka)}e^{\upi n\theta_j}.
\end{align}

We solve for the unknown coefficients $C_n^{[j]}$ using the self-consistent theory of multiple scattering, which we briefly outline here. The reader is directed to \cite{martin2006} for further details. First, we use Graf's addition theorem for Bessel functions (theorem 2.11 in \cite{martin2006}) to express the incident wave in the local polar coordinates of each scatterer as
\begin{equation}\label{incident_potential}
J_m(kr)e^{\upi m\theta}=\sum_{n=-\infty}^\infty F_n^{[l]} \frac{J_n(kr_l)}{J_n^\prime(ka)}e^{\upi n\theta_l},
\end{equation}
where
\begin{equation}
    F_{n}^{[l]} = \exp(\upi(m-n)\varphi_{0l}) J_{m-n}(kR_{0l})J_n^\prime(ka).
\end{equation}
Second, we use Graf's addition theorem for Hankel functions (theorem 2.12 in \cite{martin2006}) to express the velocity potential of the wave scattered by the $j$th scatterer in the local polar coordinates of the $l$th scatterer, namely
\begin{equation}\label{graf}
    \breve{\phi}_{\mathrm{Sc}}^{[j]}(r_l,\theta_l) = \sum_{\nu=-\infty}^\infty \left( \sum_{n=-\infty}^\infty T_{\nu n}^{[l,j]}C_n^{[j]}\right)\frac{J_\nu(kr_l)}{J_\nu^\prime(ka)}e^{\upi\nu\theta_l},
\end{equation}
where we have defined
\begin{equation}
    T_{\nu n}^{[l,j]} = \exp(\upi(n-\nu)\varphi_{jl}) \frac{H_{n-\nu}^{(1)}(kR_{jl})J_\nu^\prime (ka)}{H_n^{(1)\prime}(ka)}.
\end{equation}
in which $(R_{jl},\varphi_{jl})$ denotes the polar coordinates of $(x_l,y_l)$ with respect to $(x_j,y_j)$. The velocity potential of the wave that is incident onto the $l$th scatterer, which is a superposition of the incident wave \eqref{incident_potential} and the waves scattered by all other scatterers $j \neq l$ \eqref{graf}, is then
\begin{equation}
    \breve{\phi}_{\mathrm{I}}^{[l]}(r_l,\theta_l) = \sum_{\nu=-\infty}^\infty A_\nu^{[l]}\frac{J_\nu(kr_l)}{J_\nu^\prime(ka)}e^{\upi n\theta_l},
\end{equation}
where
\begin{equation}\label{graf2}
    A_\nu^{[l]} = F_{\nu}^{[l]} +\sum_{\substack{j=1\\j\neq l}}^N\sum_{n=-\infty}^\infty  T_{\nu n}^{[l,j]} C_n^{[j]}.
\end{equation}
From the Neumann boundary conditions on the cylinders \eqref{Neumann_BC_FD}, we must have $C_n^{[l]}=-A_n^{[l]}$ for all $n\in\Z$ and for all $1\leq l\leq N$. Thus
\begin{equation}
    C_\nu^{[l]}+\sum_{\substack{j=1\\j\neq l}}^N\sum_{n=-\infty}^\infty  T_{\nu n}^{[l,j]} C_n^{[j]}=-F_{\nu}^{[l]}.
\end{equation}
After truncating the infinite sum to $-N_{\mathrm{sol}}\leq n\leq N_{\mathrm{sol}}$, this can be written in matrix form as
\begin{equation}\label{multiple_scattering_matrix_eq}
    \mathcal{M}(\omega)\begin{bmatrix}
        \mathbf{C}^{[1]}\\\vdots\\\mathbf{C}^{[N]}
    \end{bmatrix}=-\begin{bmatrix}
        \mathbf{F}^{[1]}\\\vdots\\\mathbf{F}^{[N]}
    \end{bmatrix},
\end{equation}
where $\mathbf{C}^{[j]}$ and $\mathbf{F}^{[j]}$ are vectors of length $2N_{\mathrm{Sol}}+1$ with entries $C_n^{[j]}$ and $F_n^{[j]}$, respectively. The frequency-dependent matrix introduced in \eqref{multiple_scattering_matrix_eq} is defined as
\begin{equation}
    \mathcal{M}(\omega) = \begin{bmatrix}
    I&T^{[1,2]}&\dots&T^{[1,N]}\\
    T^{[2,1]}&I&\dots&T^{[2,N]}\\
    \vdots&\vdots&&\vdots\\
    T^{[N,1]}&T^{[N,2]}&\dots&I
    \end{bmatrix},
\end{equation}
where $I$ is the $(2N_{\mathrm{sol}}+1)$-dimensional identity matrix and $T^{[l,j]}$ is the $(2N_{\mathrm{Sol}}+1)$-dimensional square matrix with entries $T_{\nu n}^{[l,j]}$. Solving \eqref{multiple_scattering_matrix_eq} yields the unknown coefficients of the scattered waves. In turn, these can be substituted into \eqref{multipole1} in order to compute the total potential $\breve{\phi}_m(\bx,\omega)$ as required.

To generate the results in \textsection\ref{results_sec}, we choose $N_{\mathrm{sol}}=30$. This parameter was chosen to be sufficiently large so that at the points $\bx=(x,y)=(0,0)$ and $\bx=(1,0)$, $\breve{\phi}_m(\bx,\omega)$ has converged to 14 decimal places for all $|m|<30$, when sampled at 80 frequency points spanning $\omega\in[0.1,20]$.

\section{The discrete GEM}\label{discrete_GEM_sec}
Let $\mathscr{D}\subset\Omega$ be a bounded region and let $\{\bx_j|1\leq j\leq N_\bx\}$ be a set of points in $\mathscr{D}$. To approximate the integral in \eqref{GEM_spectrum} numerically, we first truncate the integral over $\Omega$ to an integral over $\mathscr{D}$, then apply a quadrature rule, i.e.:
\begin{align}
    \int_{\Omega} h(\bx)\upd\bx\approx\int_\mathscr{D} h(\bx)\upd\bx\approx \sum_{j=1}^{N_\bx} h(\bx_j) \mathrm{w}_j\label{quadrature_rule}
\end{align}
where $\mathrm{w}_j$ are the quadrature weights. The truncated spatial domain $\mathscr{D}$ should contain all points where $[f,\upi g]^\intercal(x)$ is non-negligible. If the scatterer has no area (such as in the case of the SRR), it is reasonable to choose the points $\{\bx_j\}$ to be the midpoints of a rectangular grid over $\mathscr{D}$. If the horizontal and vertical spacing  of the grid are $\Delta x$ and $\Delta y$, respectively, then the midpoint quadrature rule is implemented by setting $\mathrm{w}_j=\Delta x\,\Delta y$ for all $j$. In more general cases where the scatterer has non-zero area (such as in the case of the array of cylinders), $\{\bx_j\}$ can be taken to be the nodes of a triangulation of $\mathscr{D}$. To find the quadrature weights $\mathrm{w}_j$, we use the corner quadrature rule, which approximates the integral over each triangle as the average of the function value at the corners of the triangle multiplied by the area of the triangle \cite{larson2013}. This rule is an extension of the trapezoidal quadrature rule in one dimension.

When approximated using a quadrature rule of the form \eqref{quadrature_rule}, \eqref{GEM_spectrum} becomes
\begin{equation}\label{discrete_GEM_1}
    A_m(\omega)\approx\frac{1}{4\pi c^2}
   \sum_{j=1}^{N_\bx} \left(\upi g(\bx_j)+\omega f(\bx_j)\right)\breve{\phi}_m^*(\bx_j,\omega) \mathrm{w}_j.
\end{equation}
Consider the following discretisation of $(0,\omega_{\mathrm{max}}]$:
\begin{equation}
    \{\omega_l|1\leq l \leq N_\omega\}\label{frequency_discretisation}
\end{equation}
where $\omega_{\mathrm{max}}$ is the truncation parameter of the frequency domain and $N_{\omega}$ is the number of frequency points. Moreover, let $\mathbf{A}_m$ be the vector with entries $A_m(\omega_l)$, $l=1,\ldots,N_{\omega}$, and $\mathbf{f}$ and $\mathbf{g}$ be vectors with entries $f(\bx_j)$ and $g(\bx_j)$, respectively. Then \eqref{discrete_GEM_1} can be written in terms of matrix multiplication as
\begin{equation}
    \mathbf{A}_m=\frac{1}{4\pi c^2}\left(\upi \breve{\Phi}_m^\dag\mathcal{W}_{\bx} \mathbf{g}+\mathrm{Diag}(\omega_l)\breve{\Phi}_m^\dag\mathcal{W}_{\bx} \mathbf{f}\right),
\end{equation}
where $\breve{\Phi}_m$ is the matrix with entries $\phi_m(\bx_j,\omega_l)$ and $\mathcal{W}_{\bx}$ is the diagonal matrix with entries $\mathrm{w}_j$.

The following quadrature rule is used to approximate the time-domain solution:
\begin{equation}
    \int_0^\infty h(\omega)\upd\omega\approx\int_0^{\omega_{\mathrm{max}}} h(\omega)\upd\omega\approx\sum_{l=1}^{N_\omega}h(\omega_l)\Delta\omega_l,
\end{equation}
where $\Delta\omega_l$ is the width of the subinterval of the discretisation \eqref{frequency_discretisation} containing $\omega_l$. Using this quadrature rule, the time-domain solution given in \eqref{FD_expansion2} can then be approximated as
\begin{equation}\label{Discrete_GEM_matrix}
    \boldsymbol{\phi}_t=2\,\Re\left\{ \sum_{m=-N_\mathrm{inc}}^{N_\mathrm{inc}} \breve{\Phi}_m \mathcal{W}_\omega E_t \mathbf{A}_m\right\},
\end{equation}
where $\boldsymbol{\phi}_t$ is the vector with entries $\phi(\bx_j,t)$, $\mathcal{W}_\omega$ is the diagonal matrix with diagonal entries $\Delta\omega_l$ and $E_t$ is the diagonal matrix with entries $e^{-\upi\omega_l t}$. Note that we have also truncated the infinite sum over the angular modes. Equation \eqref{Discrete_GEM_matrix} can be equivalently stated as the following block matrix product:
\begin{equation}
    \boldsymbol{\phi}_t=2\,\Re\left\{ [\breve{\Phi}_{-N_{\mathrm{Inc}}} \dots \breve{\Phi}_{N_{\mathrm{Inc}}} ] \begin{bmatrix}
        \mathcal{W}_\omega E_t&&\\
        &\ddots&\\
        &&\mathcal{W}_\omega E_t
    \end{bmatrix}\begin{bmatrix}
        \mathbf{A}_{-N_\mathrm{sol}}\\
        \vdots\\
        \mathbf{A}_{N_\mathrm{sol}}
    \end{bmatrix}\right\}.
\end{equation}
The representation of the discrete GEM in terms of matrix multiplication was introduced in \cite{Meylan2023MWSW03}.

The most computationally expensive step of the discrete GEM is the construction of the matrices $\breve{\Phi}_{m}$. With this in mind, it is beneficial to choose $N_\omega$ and $N_{\mathrm{Inc}}$ to be as small as possible, as long as the initial conditions are accurately approximated. The parameter $N_{\mathrm{Inc}}$ must be large enough to resolve the angular variability of the initial conditions. The frequency points $\{\omega_l\}$ must be sufficiently dense in order to resolve any strong resonances. The task of optimally choosing these points is an open question, but we present a heuristic in \textsection \ref{results_sec} which assumes that the complex resonant frequencies are known.

\section{The singularity expansion method}\label{SEM_sec}
\subsection{Derivation}
We define the Fourier transform and its inverse as
\begin{equation}\label{fourier_transform}
    \widehat{h}(\omega) = \int_0^\infty h(t)e^{\upi\omega t}\upd t\quad\text{and}\quad h(t)=\frac{1}{2\pi}\intinf \widehat{h}(\omega)e^{-\upi\omega t}\upd \omega,
\end{equation}
respectively, where $h(t)$ is assumed to vanish for $t<0$. This is equivalent to the Laplace transform with the change of variables $s=-\upi\omega$. If $h$ is a real valued function, it follows from \eqref{fourier_transform} that $\widehat{h}(\omega)^*=\widehat{h}(-\omega)$ and in such cases, the inverse transform can be written as
\begin{equation}
    h(t)=\frac{1}{\pi}\Re\left\{\int_0^\infty \widehat{h}(\omega)e^{-\upi\omega t}\upd \omega\right\}.
\end{equation}
Taking the Fourier transform of \eqref{two_component_TD} gives
\begin{equation}\label{inhomogeneous_helmholtz}
    \upi (\mathscr{P}-\omega)\begin{bmatrix}
        \widehat{\phi}\\ \upi \widehat{\eta}
    \end{bmatrix}(\bx,\omega)=\begin{bmatrix}
        f\\ \upi g
    \end{bmatrix}(\bx).
\end{equation}
The time-domain solution can therefore be expressed by inverting the operator $\mathscr{P}-\omega$, then applying the inverse Fourier transform. We obtain
\begin{equation}
    \phi(\bx,t)=\Re\left\{
    -\frac{\upi}{\pi}[1,0]\int_0^\infty (\mathscr{P}-\omega)^{-1}\begin{bmatrix}
        f\\ \upi g
    \end{bmatrix}(\bx)\upd\omega\right\},
\end{equation}
since $\phi$ is real, where multiplication on the left by $[1,0]$ denotes taking the first component of the two-component vector function. In order to derive the SEM approximation formula, we first approximate this integral as
\begin{equation}\label{td_approx_contour}
    \phi(\bx,t)\approx\Re\left\{-\frac{\upi}{\pi}[1,0]\lim_{R\to\infty}\int_{\gamma_R} (\mathscr{P}-\omega)^{-1}\begin{bmatrix}
        f\\ \upi g
    \end{bmatrix}(\bx)\upd\omega\right\},
\end{equation}
where $\gamma_R$ is the negatively oriented contour that traverses the boundary of the quarter circle given by
\begin{equation}
    \{\omega:|\omega|<R,\Im(\omega)<0<\Re(\omega)\}.\nonumber
\end{equation}
In other words, this approximation assumes that the integral receives no contribution from the non-real parts of this contour. After applying the residue theorem to \eqref{td_approx_contour} and the Steinberg's formula for the residues \cite{steinberg1968meromorphic}, \eqref{td_approx_contour} then gives rise to the SEM approximation to the time-domain solution, which is
\begin{equation}\label{SEM_formula}
    \phi(r,\theta,t)\approx2\Re\left\{\sum_{j\in\mathscr{E}} \frac{\left\langle\begin{bmatrix}
        f\\\upi g
    \end{bmatrix},\begin{bmatrix}
        \psi_j\\\upi\zeta_j
    \end{bmatrix}\right\rangle}{\left\langle\begin{bmatrix}
        \phi_j\\\upi\eta_j
    \end{bmatrix},\begin{bmatrix}
        \psi_j\\\upi\zeta_j
    \end{bmatrix}\right\rangle}\phi_j(r,\theta)e^{-\upi\omega_j^\circ t},\right\}
\end{equation}
where $\mathscr{E}$ indexes all of the complex resonances $[\phi_j,\upi\eta_j]^\intercal$, which are eigenfunctions of $\mathscr{P}$ associated with the eigenvalue $\omega_j^\circ$. In particular, they satisfy
\begin{subequations}\label{resonance_problem}
\begin{equation}
    \mathscr{P}\begin{bmatrix}
        \phi_j\\\upi\eta_j
    \end{bmatrix}=\omega_j^\circ\begin{bmatrix}
        \phi_j\\\upi\eta_j
    \end{bmatrix},
\end{equation}
and the Sommerfeld radiation condition
\begin{equation}
    \sqrt{r}(\partial_r-\upi k_j^\circ)\phi_j\to 0\quad\text{ as }r\to\infty,
\end{equation}
\end{subequations}
where $k_j^\circ=\omega_j^\circ/c$. They can be interpreted as non-trivial solutions to the unforced frequency-domain problem, which only exist at the complex resonant frequencies $\omega_j^\circ$, and represent outgoing waves only with no incoming component. In contrast, the absorbing modes $[\psi_j,\zeta_j]^\intercal$ are non-trivial solutions to the unforced frequency-domain problem at $\omega_j^{\circ *}$, and represent incoming waves with no outgoing component. That is, they satisfy
\begin{subequations}\label{absorbing_mode_problem}
\begin{equation}
    \mathscr{P}\begin{bmatrix}
        \psi_j\\\upi\zeta_j
    \end{bmatrix}=\omega_j^{\circ *}\begin{bmatrix}
        \psi_j\\\upi\zeta_j
    \end{bmatrix},
\end{equation}
as well as the adjoint of the Sommerfeld radiation condition
\begin{equation}
    \sqrt{r}(\partial_r+\upi k_j^{\circ *})\psi_j\to 0\quad\text{ as }r\to\infty.
\end{equation}
\end{subequations}
It is straightforward to see that if $[\phi_j,\upi\eta_j]^\intercal$ satisfies \eqref{resonance_problem}, then $[\phi_j^*,-\upi\eta_j^*]^\intercal$ satisfies $\eqref{absorbing_mode_problem}$, thus we take $\psi_j=\phi_j^*$ and $\zeta_j=-\eta_j^*$ in what follows.

Note that in problems with symmetry, the complex resonant frequencies $\omega_j^\circ$ may be poles of $(\mathscr{P}-\omega)^{-1}[f,\upi g]^\intercal$ of order greater than one \cite{meylan2009time,WOLGAMOT2017232}. In these cases, the eigenspace of a complex resonant frequency $\omega_j^\circ$ can be the span of two or more linearly independent complex resonant modes. However, this is already accounted for in the SEM approximation formula \eqref{SEM_formula} because $\mathscr{E}$ indexes over the the modes and not the resonant frequencies themselves (i.e.\ it is possible that $\omega_j=\omega_l$ for $j\neq l$).

\subsection{Finding the resonant frequencies and modes}
Since they are single frequency solutions with no incoming component, the complex resonant frequencies and resonant modes can be numerically approximated from the nonlinear eigenvalue problem
\begin{equation}\label{classic_matrix_nullspace}
    \mathcal{M}(\omega_j)\mathbf{\widetilde{x}}=\mathbf{0},
\end{equation}
This should be contrasted with \eqref{classic_matrix_eq}, which is the matrix equation that underlies the frequency domain scattering problem and has a non-zero right hand side due to the presence of incoming waves. Thus the complex resonant frequencies $\omega_j$ are precisely those at which $\mathcal{M}(\omega_j)$ has a non-trivial kernel, or equivalently, those at which $\det\mathcal{M}(\omega_j)=0$ (up to numerical error). The null vector $\mathbf{\widetilde{x}}$ then determines the resonant mode. In the case of the SRR, \eqref{recover_C_and_D} can be used to recover the coefficients of the mode from $\mathbf{\widetilde{x}}$, whereas in the case of the array of cylinders, \eqref{multiple_scattering_matrix_eq} shows that $\mathbf{\widetilde{x}}$ already contains the unknown coefficients. The modes obtained using this procedure are numerical solutions to \eqref{resonance_problem}, that is, they satisfy the boundary conditions and the Sommerfeld radiation condition in the absence of an incident wave.

\begin{figure}
    \centering
    \includegraphics[width=0.75\textwidth]{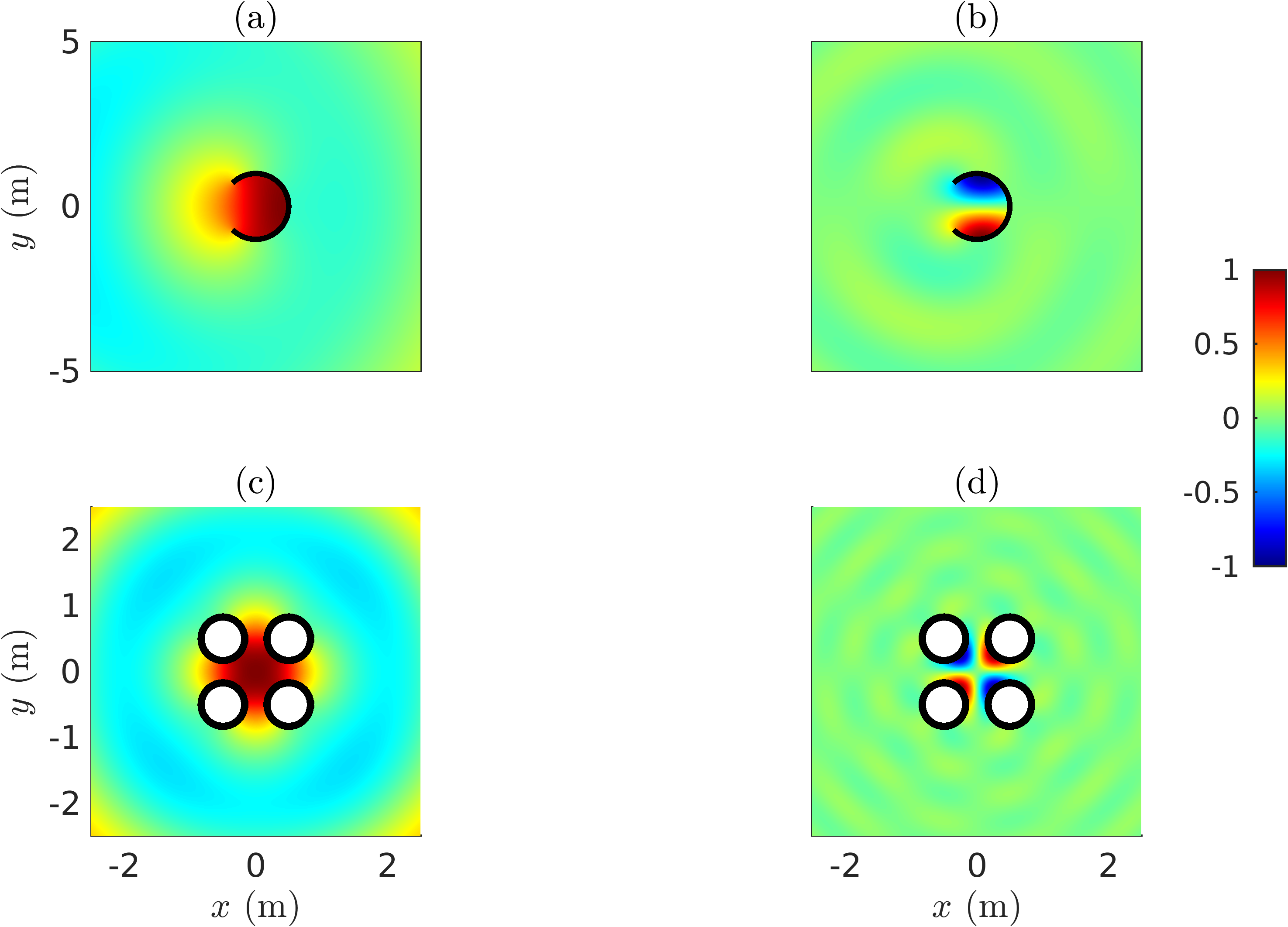}
    \caption{(a,b) Resonant modes of a SRR with parameters $a=1$\,m, $\alpha=\pi/4$ and $\beta=\pi$, at the complex resonant frequency (a) $(0.5923-0.1126\upi)$\,s$^{-1}$ and (b) 
   $(1.9437-0.0294\upi)$\,s$^{-1}$. (c,d) Resonant modes of an array of four cylinders with radius $a=0.33$\,m and with centre-points $(0.5,0.5)$\,m, $(-0.5,0.5)$\,m, $(0.5,-0.5)$\,m and $(-0.5,-0.5)$\,m, at the complex resonant frequencies (c) $(1.7648-0.3372\upi)$\,s$^{-1}$ and (d) $8.2132-0.0410\upi)$\,s$^{-1}$. In all panels, the colours indicate the values of $\Re(\phi_j(\bx))$.
}
\label{fig:resonant_modes}
\end{figure}

A major challenge for implementing the SEM is the task of finding these complex resonances numerically. This should be done carefully, because omitting the contribution from a resonance that is strongly excited by the initial data $[f,\upi g]^\intercal$ would lead to inaccurate outputs. Thus, both efficiency and robustness are important criteria when selecting an algorithm. Our method, which uses (i) a global search followed by (ii) a local refinement, is as follows:
\begin{enumerate}[label=(\roman*)]
    \item First, we apply a global root-finding algorithm, which uses Cauchy's argument principle and the assumption that $\det\mathcal{M}(\omega)$ is analytic. Such algorithms are advantageous because they attempt to find all of the roots within a given region and their multiplicities. We use the global complex roots and poles finding algorithm of Kowalczyk \cite{kowalczyk2015complex,Kowalczyk2018}. This method uses a discrete form of Cauchy's argument principle and an iteratively-refined triangular mesh of a region of the complex plane to find the zeros within that region. Other examples of global root-finding algorithms include the algorithm of Delves and Lyness \cite{delves1967numerical}, which uses Cauchy's argument principle to convert the nonlinear root-finding problem into a system of polynomial equations, and that of Meylan and Gross \cite{meylan2002parallel}, which generalises the bisection method to analytic functions. Because global root-finding algorithms typically suffer from slow convergence, we run Kowalczyk's algorithm at a high tolerance, as the numerical approximations of the roots can be refined more efficiently using step (ii).
    \item Second, we use a local root-finding algorithm to refine the initial guesses from the global algorithm. Here, we apply the method of Wolgamot et al. iteratively \cite{WOLGAMOT2017232}. This method advances towards a root by solving a generalised eigenvalue problem in terms of the matrix derivative of $\mathcal{M}$. The matrix derivative is approximated by the finite difference formula $\mathcal{M}^\prime(\omega)\approx(\mathcal{M}(\omega+\delta)-\mathcal{M}(\omega))/\delta$, where we use $\delta=10^{-7}$. From a suitable initial guess, we typically only require three iterations to find an eigenvalue-eigenvector pair $(\omega_j,\mathbf{v}_j)$ that satisfies
    \begin{equation}
    \frac{\|\mathcal{M}(\omega_j)\mathbf{v}_j\|}{\|\mathbf{v}_j\|}<10^{-11}.
    \end{equation}
    This criteria ensures that the complex resonances used in the SEM accurately satisfy \eqref{resonance_problem}. Although iterative methods are typically much more efficient than global methods, we do not use them initially because they do not offer confidence that all of the roots have been found.
\end{enumerate}
Figure \ref{fig:resonant_modes} shows a collection of resonant modes of a SRR and an array of four cylinders.

\subsection{Evaluating the normalisation coefficients}\label{normalisation_coeffs_subsec}
Let us now compute the denominator of the terms in the SEM expansion \eqref{SEM_formula}, i.e.
\begin{subequations}
    \begin{align}
\langle\boldsymbol{\phi}_j,\boldsymbol{\psi}_j\rangle&=\left\langle\begin{bmatrix}
        \phi_j\\\upi\eta_j
    \end{bmatrix},\begin{bmatrix}
        \psi_j\\\upi\zeta_j
    \end{bmatrix}\right\rangle\\
    &=\int_{\Omega}\frac{1}{c^2}\eta_j\zeta_j^*+\nabla\phi_j\cdot\nabla\psi_j^*\upd\bx\\
    &=I_{\Omega\setminus\mathscr{A}}+I_{\mathscr{A}},
\end{align}
\end{subequations}
where $I_{\Omega\setminus\mathscr{A}}$ and $I_\mathscr{A}$ are the values of the integral over the regions $\Omega\setminus\mathscr{A}$ and $\mathscr{A}$, respectively, where the annular region $\mathscr{A}$ is given by
\begin{equation*}
    \mathscr{A}=\{(r,\theta)\in\Omega|r\in(b,\infty)\}.
\end{equation*}
The radius $b$ is chosen to be sufficiently large so that it encloses the scatterer, i.e.\ $\Gamma\cap\mathscr{A}=\varnothing$.

In the outer region $\mathscr{A}$, we expand the resonant and absorbing modes as
\begin{subequations}\label{resonant_absorbing_modes}
   \begin{align}
    \phi_j(r,\theta) &= \sum_{n=-\infty}^\infty \widetilde{C}_n^{(j)}H_n^{(1)}(k_jr)e^{\upi n \theta},\label{resonant_mode}\\
    \psi_j(r,\theta) &= \sum_{n=-\infty}^\infty(-1)^n \widetilde{C}_{-n}^{(j)*}H_n^{(2)}(k_j^*r)e^{\upi n \theta},\label{absorbing_mode}
\end{align} 
\end{subequations}
respectively, where \eqref{absorbing_mode} follows from the facts that $\psi_j=\phi_j^*$, $H_n^{(1)}(z)^*=H_n^{(2)}(z^*)$ and $H_{-n}^{(2)}(z)=(-1)^n H_{n}^{(2)}(z)$ \cite{abramowitz1988handbook}. With reference to \eqref{SRR_scattered}, the resonant modes of the SRR are already in the form of \eqref{resonant_mode}, where $\widetilde{C}_n^{(j)}=C_n^{(j)}/H_n^{(1)}(k_ja)$ and $C_n^{(j)}$ are the coefficients of the $j$th resonant mode. The resonant modes of the array of cylinders are not in the form of \eqref{resonant_mode}, but they can be converted into this form by invoking Graf's addition theorem for Hankel functions (theorem 2.12 in \cite{martin2006}), which gives
\begin{equation}\label{exterior_expansion}
    \widetilde{C}_\nu^{(j)}=\sum_{l=1}^N\sum_{n=-\infty}^\infty (-1)^{n-\nu}\exp(\upi(n-\nu)\varphi_{0l})\frac{J_{n-\nu}(k_jR_{0l})}{H_n^{(1)\prime}(k_ja)} C_n^{[l](j)},
\end{equation}
for $r>b>\max_{j=1,\dots,N}R_{0j}$, where $C_n^{[l](j)}$ denotes the coefficients of the component of the $j$th resonant mode that radiates away from the $l$th cylinder.

\subsubsection{The outer integral}\label{outer_integral_subsec}
The integral $I_\mathscr{A}$ is divergent because $k_j=\omega_j/c$ is in the lower half of the complex plane, which causes $H_n^{(1)}(k_jr)$ and $H_n^{(2)}(k_j^*r)$ to grow exponentially with $r$---this can be seen from their asymptotic formulae for large argument \cite{abramowitz1988handbook}
\begin{subequations}\label{large_argument_hankel}
    \begin{align}
    H_n^{(1)}(z)&\sim\sqrt{\frac{2}{\pi z}}\exp\left(\upi\left(z-\frac{n\pi}{2}-\frac{\pi}{4}\right)\right)\\
    H_n^{(2)}(z)&\sim\sqrt{\frac{2}{\pi z}}\exp\left(-\upi\left(z-\frac{n\pi}{2}-\frac{\pi}{4}\right)\right).
\end{align}
\end{subequations}
However, the required value of the inner product for the SEM can be found by regularising $I_{\mathscr{A}}$, that is, by evaluating $I_{\mathscr{A}}$ as the analytic continuation of an expression which would converge if $k_j$ were in the upper half plane. First, we use the facts that $\eta_j=-\upi\omega_j\phi_j$ and $\zeta_j=-\upi\omega\psi_j$ and apply Green's first identity to obtain
\begin{align}
I_\mathscr{A}=\int_{\mathscr{A}}\left(\frac{\omega_j^2}{c^2}\phi_j\psi_j^*-\psi_j^*\bigtriangleup\phi_j\right)\upd\bx+\int_{\partial\mathscr{A}}\psi_j^*\partial_n\phi_j\upd s,
\end{align}
where $\partial_n$ is the directional derivative with respect to the outwards pointing unit normal of $\mathscr{A}$. Since $\phi_j$ satisfies the Helmholtz equation with wavenumber $k_j$, we can combine the terms in parentheses to obtain
\begin{align}
I_\mathscr{A}=2k_j^2\int_{\mathscr{A}}\phi_j\psi_j^*\upd\bx+\int_{\partial\mathscr{A}}\psi^*\partial_n\phi\upd s.\label{annular_integral_greens}
\end{align}
Using \eqref{resonant_absorbing_modes}, we compute
\begin{align}
\int_{\mathscr{A}}\phi_j\psi_j^*\upd\bx&=\int_{-\pi}^\pi \int_b^\infty \left(\sum_{n=-\infty}^\infty \widetilde{C}_n^{(j)}H_n^{(1)}(k_jr)e^{\upi n \theta}\right)\left(\sum_{l=-\infty}^\infty (-1)^m\widetilde{C}_{-l}^{(j)*}H_l^{(2)}(k_j^*r)e^{\upi l \theta}\right)^*r\upd r\upd\theta\nonumber\\
&=2\pi\sum_{n=-\infty}^\infty (-1)^n\widetilde{C}_n^{(j)}\widetilde{C}_{-n}^{(j)}\int_b^\infty H_n^{(1)}(k_jr)^2r\upd r,\label{Hankel_integral}
\end{align}
where we have used the orthogonality of the Fourier terms $e^{\upi n\theta}$ and the fact that $H_n^{(2)^*}(z^*)=H_n^{(1)}(z)$ \cite{abramowitz1988handbook}. To compute the integral appearing in \eqref{Hankel_integral}, we use the following formula for the indefinite integral due to Watson (equation 5.11.11 in \cite{watson1922treatise}):
\begin{equation}\label{watson_integral}
    \int \mathscr{C}_n(\kappa r)^2r\upd r=\tfrac{1}{2}r^2(\mathscr{C}_n(\kappa r)^2-\mathscr{C}_{n-1}(\kappa r)\mathscr{C}_{n+1}(\kappa r)),
\end{equation}
where $\mathscr{C}_n$ is any cylinder function of order $n$ (in particular, \eqref{watson_integral} holds for $\mathscr{C}_n=J_n$ or $\mathscr{C}_n=H_n^{(1)}$). The corresponding definite integral over $(b,\infty)$ converges for $\Im(\kappa)>0$ and is equal to
\begin{equation}
    \int_b^\infty H_n^{(1)}(\kappa r)^2r\upd r=\tfrac{1}{2}b^2(H_{n-1}^{(1)}(\kappa b)H_{n+1}^{(1)}(\kappa b)-H_n^{(1)}(\kappa b)^2).\label{watson_integral_definite}
\end{equation}
The asymptotic formulae for Hankel functions at large arguments \eqref{large_argument_hankel} were used to eliminate the term from the upper bound of the integral. The right hand side of \eqref{watson_integral_definite} can be analytically continued to $\Im(\kappa)<0$ (i.e.\ to the region containing the complex resonant frequencies $k_j$, which allows us to regularise \eqref{Hankel_integral} as
\begin{equation}
\int_{\mathscr{A}}\phi_j\psi_j^*\upd\bx=\pi b^2\sum_{n=-\infty}^\infty (-1)^n\widetilde{C}_n^{(j)}\widetilde{C}_{-n}^{(j)}(H_{n-1}^{(1)}(k_j b)H_{n+1}^{(1)}(k_j b)-H_n^{(1)}(k_j b)^2).
\end{equation}
We must also regularise the boundary integral appearing in \eqref{annular_integral_greens}. We compute
\begin{align}\label{boundary_integral_unregularised}
\int_{\partial\mathscr{A}}\psi_j^*\partial_n\phi_j\upd s&=-b\int_{-\pi}^\pi \psi_j^*(b,\theta)\partial_r\phi_j(b,\theta)\upd \theta+\lim_{R\to\infty}R\int_{-\pi}^\pi \psi_j^*(R,\theta)\partial_r\phi_j(R,\theta)\upd \theta,
\end{align}
in which the limit diverges. Nevertheless, with the intention of regularising the limit, we evaluate the integral to obtain
\begin{align}
    R\int_{-\pi}^\pi \psi_j^*(R,\theta)\partial_r\phi_j(R,\theta)\upd \theta=2\pi k_j R\sum_{n=-\infty}^\infty (-1)^n\widetilde{C}_n^{(j)}\widetilde{C}_{-n}^{(j)}H_n^{(1)}(k_jR)H_n^{(1)\prime}(k_jR).
\end{align}
From \eqref{large_argument_hankel}, we have
\begin{equation}
    \lim_{R\to\infty}R H_n^{(1)}(\kappa R)H_n^{(1)\prime}(\kappa R)=0,
\end{equation}
for all $\Im(\kappa)>0$. The analytic continuation of this limit to $\Im(\kappa)<0$ is trivially zero. Therefore we can write the regularised form of \eqref{boundary_integral_unregularised} as
\begin{subequations}
\begin{align}
\int_{\partial\mathscr{A}}\psi_j^*\partial_n\phi_j\upd s&=-b\int_{-\pi}^\pi \psi_j^*(b,\theta)\partial_r\phi_j(b,\theta)\upd \theta\\
&=-2\pi k_j b\sum_{n=-\infty}^\infty (-1)^n\widetilde{C}_n^{(j)}\widetilde{C}_{-n}^{(j)}H_n^{(1)}(k_jb)H_n^{(1)\prime}(k_jb).
\end{align}
The complete regularised expression for $I_\mathscr{A}$ is therefore
\begin{align}
I_\mathscr{A}&=2\pi k_j^2 b^2\sum_{n=-\infty}^\infty (-1)^n\widetilde{C}_n^{(j)}\widetilde{C}_{-n}^{(j)}(H_{n-1}^{(1)}(k_j b)H_{n+1}^{(1)}(k_j b)-H_n^{(1)}(k_j b)^2)\nonumber\\
&\quad-2\pi k_j b\sum_{n=-\infty}^\infty (-1)^n\widetilde{C}_n^{(j)}\widetilde{C}_{-n}^{(j)}H_n^{(1)}(k_jb)H_n^{(1)\prime}(k_jb)\\
&=2\pi k_j b\sum_{n=-\infty}^\infty (-1)^n\widetilde{C}_n^{(j)}\widetilde{C}_{-n}^{(j)}\nonumber\\
&\quad\times\left[k_jbH_{n-1}^{(1)}(k_j b)H_{n+1}^{(1)}(k_j b)-k_jbH_n^{(1)}(k_j b)^2-H_n^{(1)}(k_jb)H_n^{(1)\prime}(k_jb)\right].\label{regularised_exterior_integral}
\end{align}
\end{subequations}
In practice, we truncate this expansion to $-N_{\mathrm{sol}}\leq n\leq N_{\mathrm{sol}}$. The regularised expression \eqref{regularised_exterior_integral} applies to a much broader range of problems than the two considered in this paper. For instance, scattering by arbitrarily shaped scatterers could be solved using the boundary element method, and the resulting solution could be expanded in the form of \eqref{resonant_mode} in the exterior domain $\mathscr{A}$.

\subsubsection{The inner integral}
It remains for us to compute $I_{\Omega\setminus\mathscr{A}}$, i.e.\ the integral over the inner region. The fact that the resonant and absorbing modes can be expanded in the form of \eqref{resonant_absorbing_modes} in the exterior domain $\mathscr{A}$, irrespective of the scatterer geometry, allowed us to obtain \eqref{regularised_exterior_integral}. An equivalent expression for the interior field $\Omega\setminus\mathscr{A}$ cannot be obtained for arbitrary scatterer geometry, thus in most cases this integral must be computed numerically. One possibility is to use the triangulation-based quadrature method described in \textsection\ref{discrete_GEM_sec}, which is how we compute $I_{\Omega\setminus\mathscr{A}}$ in the case of the array of cylinders. In analogy with the steps leading to \eqref{annular_integral_greens}, we compute
\begin{align}
I_{\Omega\setminus\mathscr{A}}=2k_j^2\int_{\Omega\setminus\mathscr{A}}\phi_j\psi_j^*\upd\bx+\int_{\partial(\Omega\setminus\mathscr{A})}\psi^*\partial_n\phi\upd s.
\end{align}
Note that the boundary integral term in the above cancels with term arising from the boundary integral in \eqref{regularised_exterior_integral} (i.e.\ the third term inside the square brackets) when $I_\mathscr{A}$ and $I_{\Omega\setminus\mathscr{A}}$ are summed because $\phi_j$ and $\partial_r\phi_j$ are both continuous across $r=b$.

In the case of the SRR, the interior domain of the resonant mode can be written as an expansion of Bessel functions of the first kind \eqref{SRR_interior}. For brevity, we rescale the coefficients to write this as
\begin{equation}
    \phi_j(r,\theta)=\sum_{n=-\infty}^\infty\widetilde{D}_n^{(j)}J_n(k_jr)e^{\upi n\theta},
\end{equation}
for $r<b$, where we have chosen $b=a$. In analogy to the derivation of \eqref{absorbing_mode}, we use the facts that $\psi_j=\phi_j^*$, $J_n(z)^*=J_n(z^*)$ and $J_{-n}(z)=(-1)^n J_{n}(z)$ \cite{abramowitz1988handbook} to obtain the absorbing mode as
\begin{equation}
    \psi_j(r,\theta)=\sum_{n=-\infty}^\infty(-1)^n\widetilde{D}_{-n}^{(j)*}J_n(k_j^*r)e^{\upi n\theta}.
\end{equation}
Following the steps in \textsection\ref{normalisation_coeffs_subsec}\ref{outer_integral_subsec}, we apply Green's first identity and Watson's formula for the definite integral \eqref{watson_integral} to obtain
\begin{align}
    I_{\Omega\setminus\mathscr{A}}&=2\pi k_j a\sum_{n=-\infty}^\infty (-1)^n\widetilde{D}_n^{(j)}\widetilde{D}_{-n}^{(j)}\nonumber\\
    &\quad\times\left[k_ja J_n(k_ja)^2-k_jaJ_{n-1}(k_ja)J_{n+1}(k_ja)+J_n(k_ja)J_n^\prime (k_ja)\right].
\end{align}

\section{Results}\label{results_sec}
As mentioned in \textsection\ref{discrete_GEM_sec}, by far the most computationally expensive step in the discrete GEM is the computation of the frequency-domain solution matrices $\breve{\Phi}_m$. To generate the results in this section, we set $N_{\mathrm{Inc}}=30$ and used $N_\omega>500$ frequency quadrature points selected using the heuristic described in the next paragraph. Computation of the full frequency-domain solution matrix $[\breve{\Phi}_{-N_{\mathrm{Inc}}},\dots,\breve{\Phi}_{N_{\mathrm{Inc}}}]$ took approximately 70 and 50 minutes for the SRR and the array of cylinders, respectively, using \textsc{MATLAB}(R) R2020a on a desktop computer running Fedora release 38 with an Intel(R) Core(TM) i5-7400 processor operating at 3.00 GHz and 48GB of RAM. The full frequency domain matrix required 33GB and 22GB of memory for the SRR and array of cylinders, respectively. Fortunately, the full frequency-domain solution matrix must only be computed once for each geometry, as it does not depend on the initial conditions. The computation time could be accelerated by reducing $N_\bx$, $N_{\mathrm{Inc}}$ or $N_\omega$, at the cost of reducing accuracy.

The following heuristic was used to select quadrature points for the discrete GEM. We begin with a uniformly-spaced discretisation of 500 points spanning $(0,\omega_{\mathrm{max}}]$. Then, for every complex resonance with Q-factor $Q_j>25$, we add 20 uniformly-spaced spanning $[\Re(\omega_j^\circ)+\Im(\omega_j^\circ),\Re(\omega_j^\circ)-\Im(\omega_j^\circ)]$. Here, the Q-factor is defined as
\begin{equation}
    Q_j\coloneqq-\frac{\Re(\omega_j)}{2\Im(\omega_j)}.
\end{equation}
This heuristic is motivated by the observations that resonances with a high Q-factor induce peaks in the spectral amplitude functions $A_m(\omega)$, and that the widths of these peaks is approximately inversely proportional to the imaginary part of the complex resonant frequency. Thus, this heuristic allows us to accurately resolve these peaks while keeping the number of frequency quadrature points $N_\omega$ relatively low.

\begin{figure}
    \centering
    \includegraphics[width=0.75\textwidth]{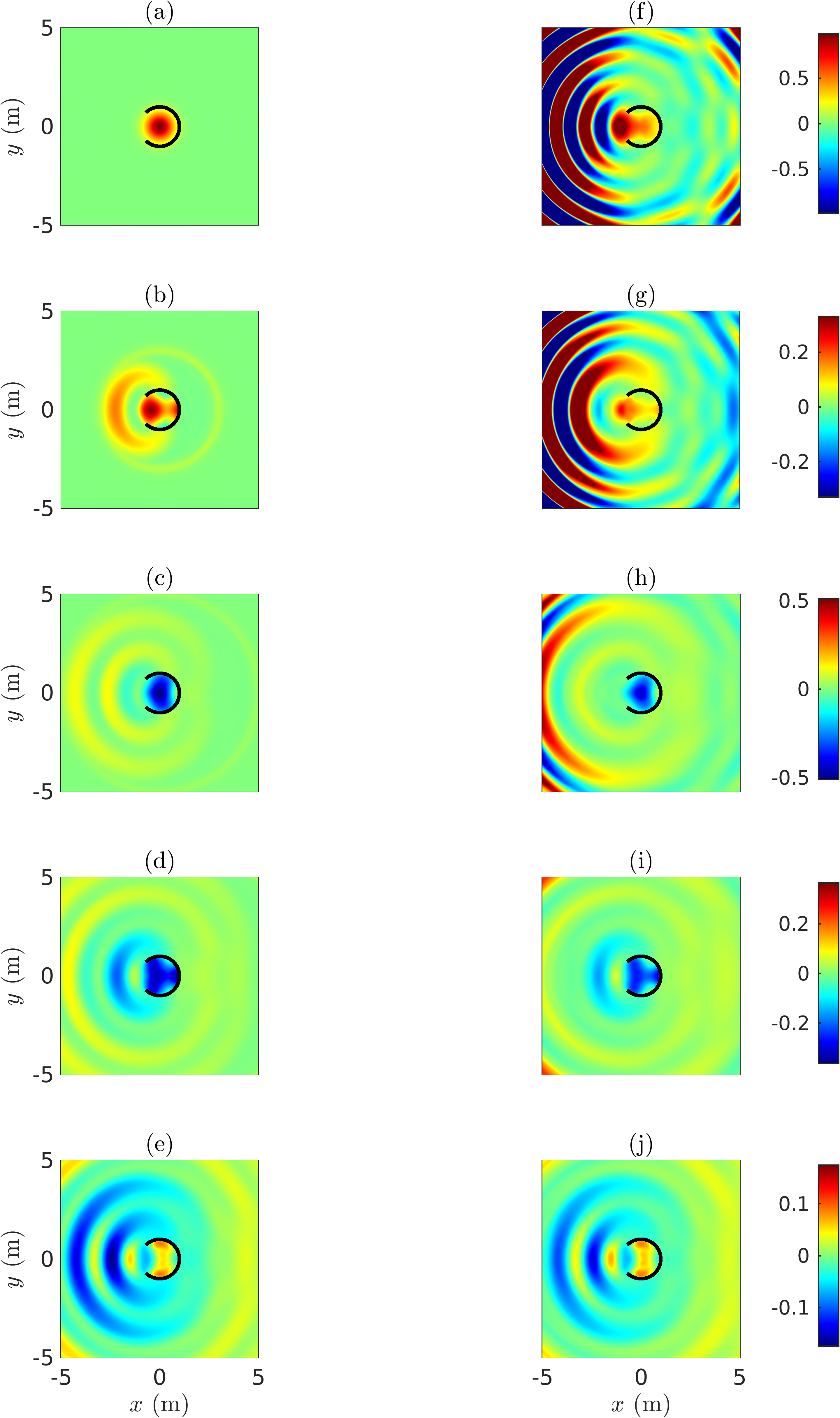}
    \caption{Temporal evolution of the velocity potential $\phi$ in the presence of the SRR with the initial condition $f(x,y)=e^{-2(x^2+y^2)}$, computed using the GEM (a--e) and the SEM (f--j). The panels correspond to the times (a,f) $t=0$\,s, (b,g) $t=2$\,s, (c,h) $t=4$\,s, (d,i) $t=6$\,s and (e,j) $t=8$\,s. The error (see \eqref{error_eq}) in the GEM solution is $0.028$. Note that the colour scales are set using the GEM solution, thus panels showing the SEM solution at the corresponding times may contain regions where the colour has saturated due to the exponential growth of the modes.}
    \label{fig:SRR_Gaussian}
\end{figure}

\begin{figure}
    \centering
    \includegraphics[width=0.75\textwidth]{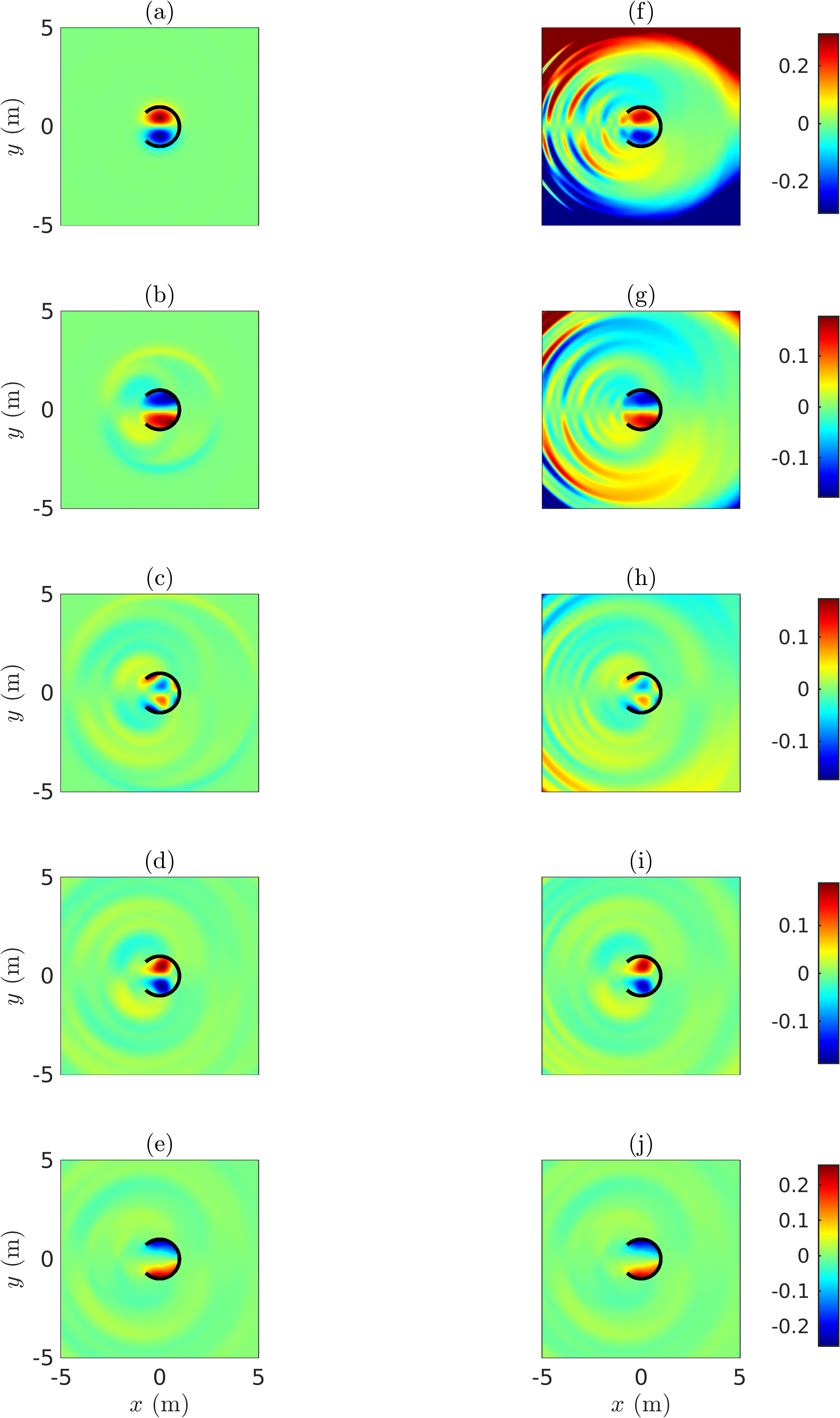}
    \caption{As for figure \ref{fig:SRR_Gaussian} with the initial disturbance $f(x,y)=\sin(y)e^{-2(x^2+y^2)}$. The error of the GEM solution is $0.029$.}
    \label{fig:SRR_dipole}
\end{figure}

\begin{figure}
    \centering
    \includegraphics[width=0.75\textwidth]{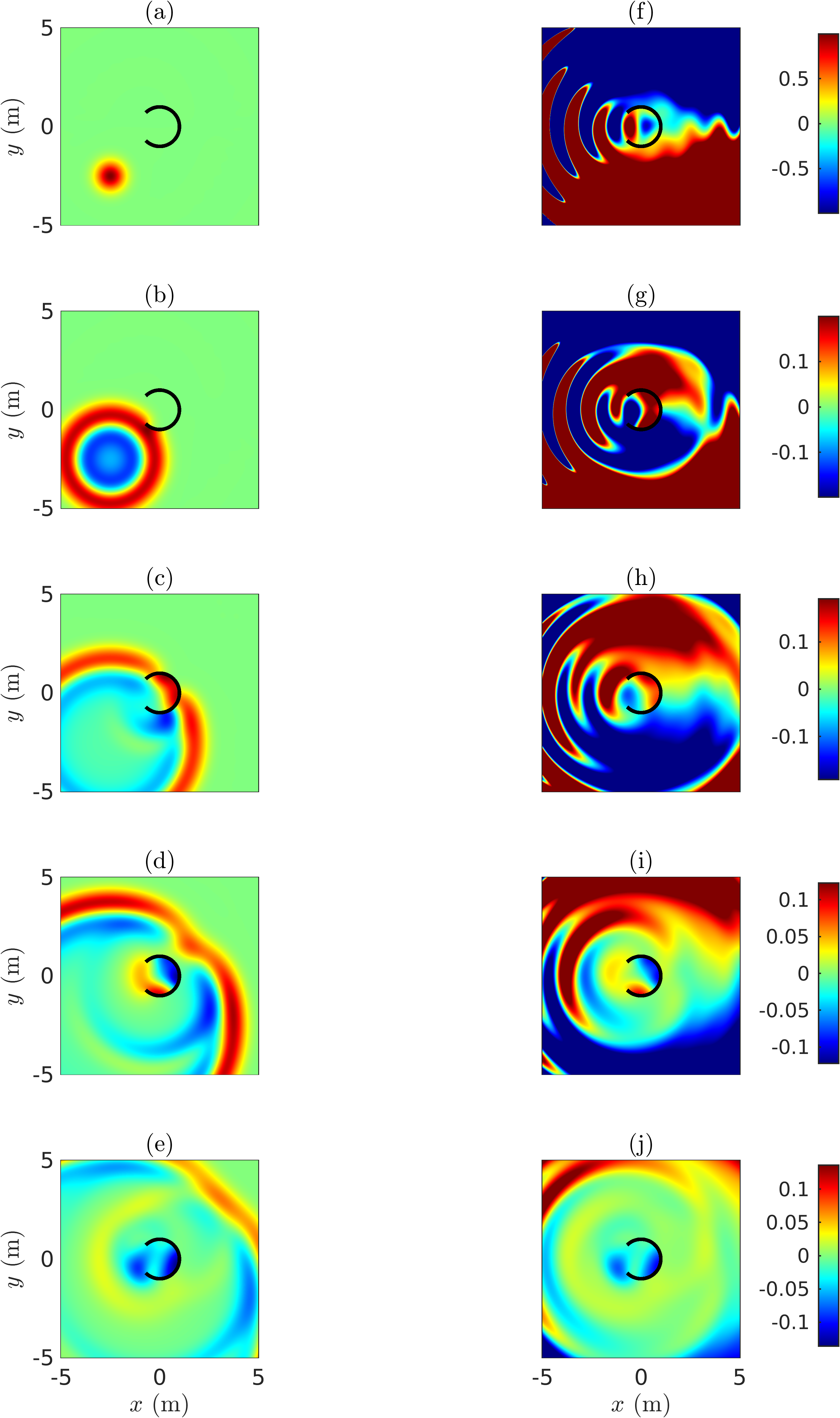}
    \caption{As for figure \ref{fig:SRR_Gaussian} with the initial disturbance $f(x,y)=e^{-2((x+2.5)^2+(y+2.5)^2)}$. The error of the GEM solution is $5.3\times10^{-4}$.}
    \label{fig:SRR_exterior_gaussian}
\end{figure}

\begin{figure}
    \centering
    \includegraphics[width=0.75\textwidth]{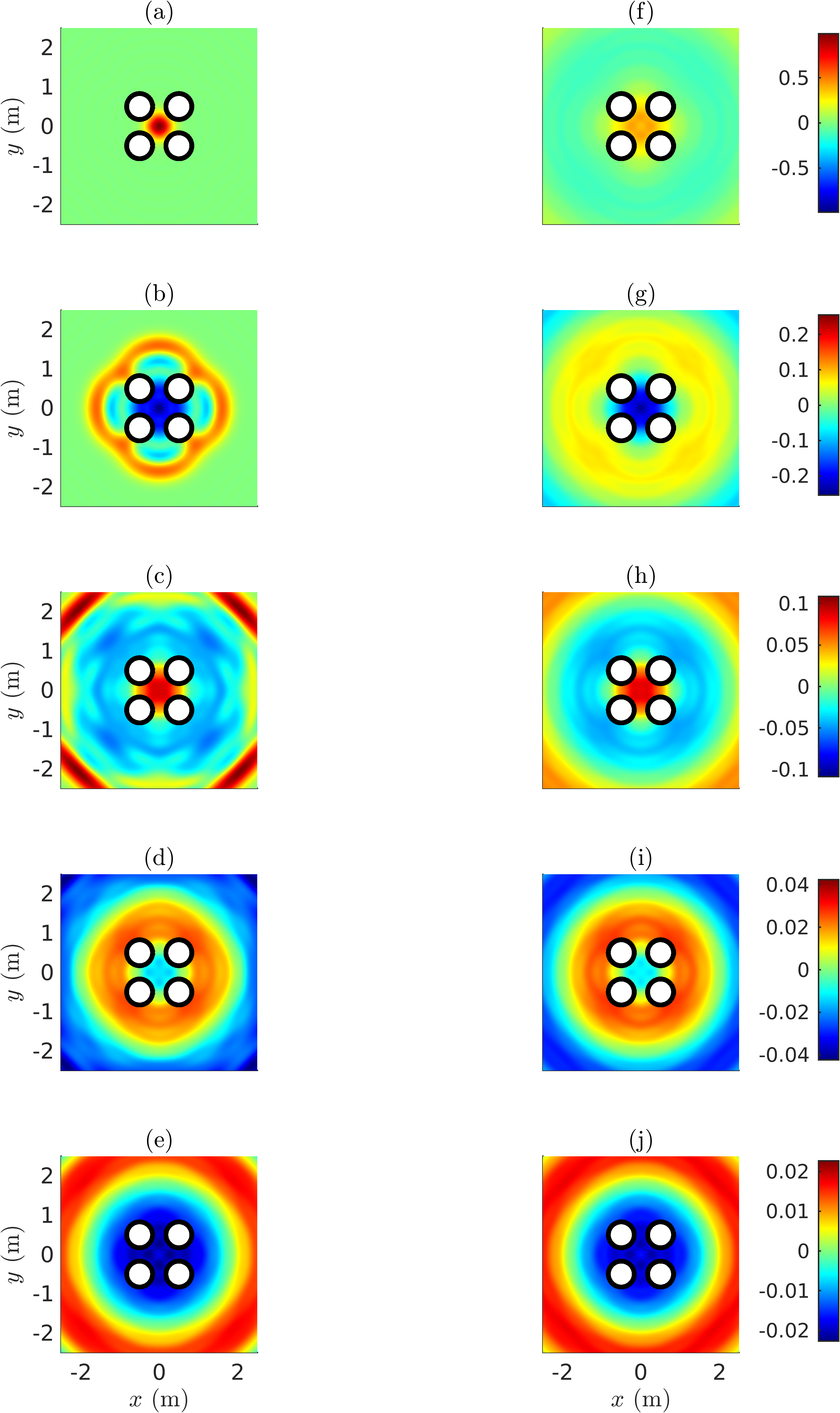}
    \caption{Temporal evolution of the velocity potential $\phi$ in the presence of the array of four cylinders with the initial condition $f(x,y)=e^{-10(x^2+y^2)}$, computed using the GEM (a--e) and the SEM (f--j). The panels correspond to the times (a,f) $t=0$\,s, (b,g) $t=1.5$\,s, (c,h) $t=3$\,s, (d,i) $t=4.5$\,s and (e,j) $t=6$\,s. The error in of the GEM solution is $0.057$.}
    \label{fig:Cylinders_interior_gaussian}
\end{figure}

\begin{figure}
    \centering
    \includegraphics[width=0.75\textwidth]{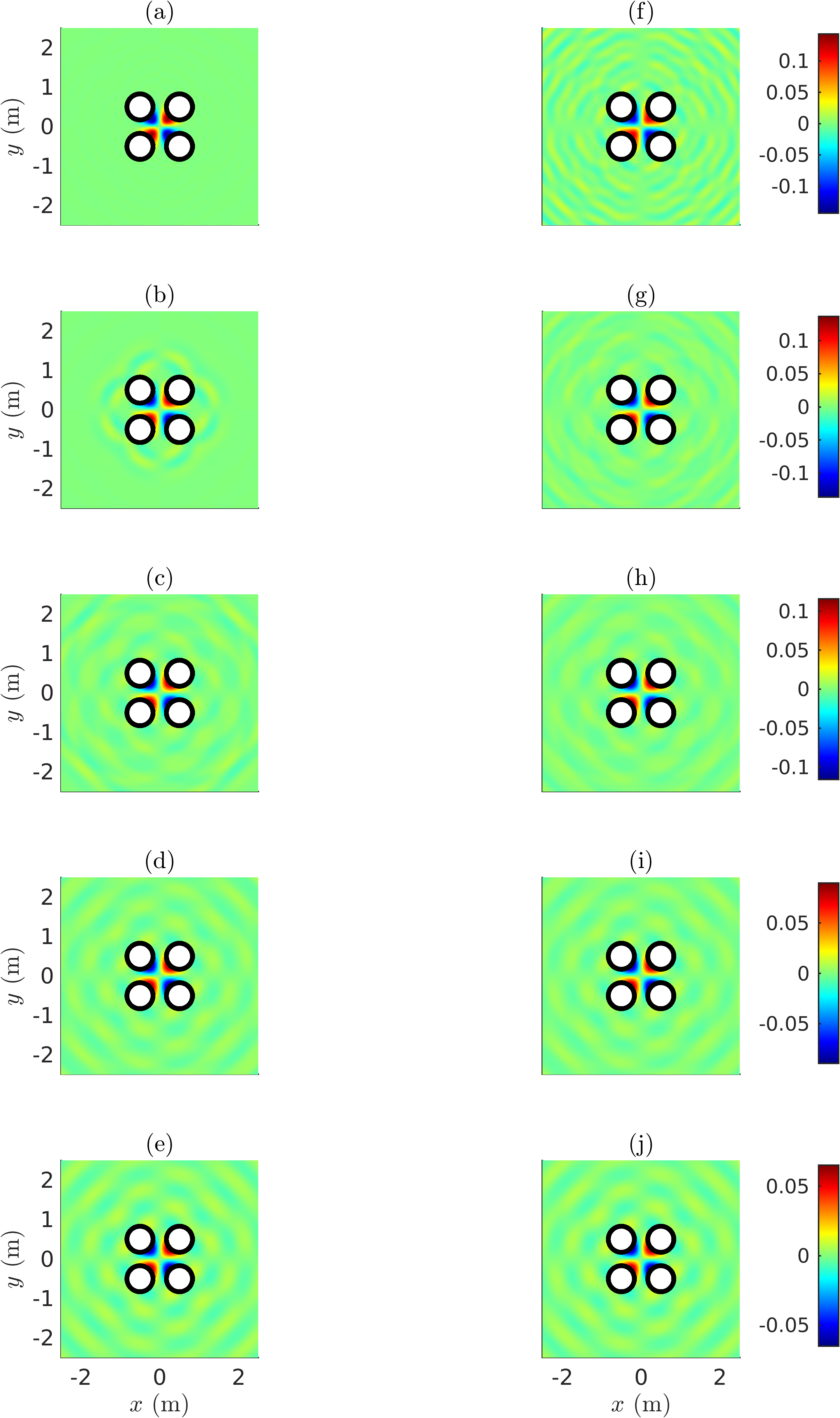}
    \caption{As for figure \ref{fig:Cylinders_interior_gaussian} with the initial disturbance $f(x,y)=\sin(3x)\sin(3y)e^{-10(x^2+y^2)}$. The error of the GEM solution is $0.018$.}
    \label{fig:Cylinders_quadrupole}
\end{figure}

\begin{figure}
    \centering
    \includegraphics[width=0.75\textwidth]{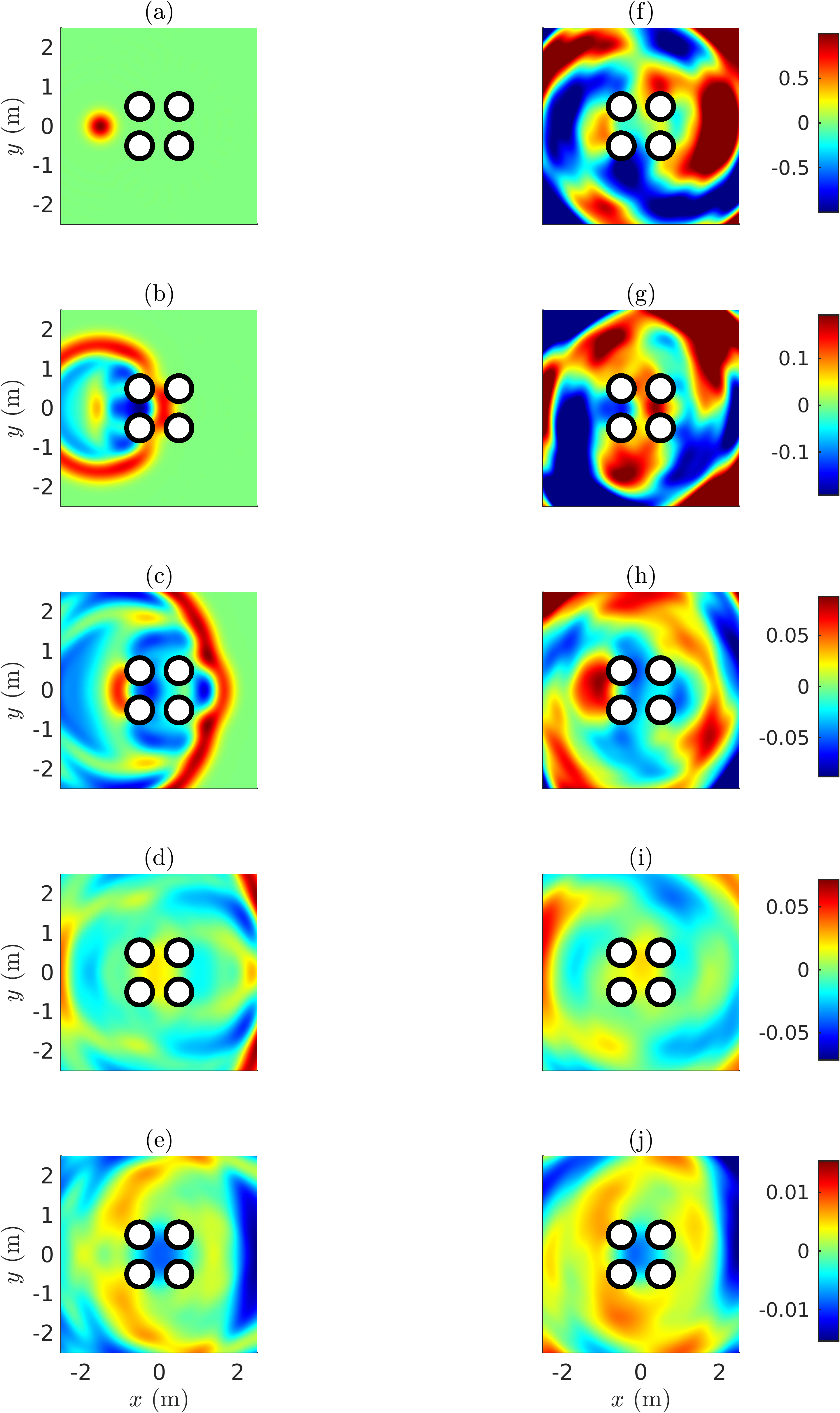}
    \caption{As for figure \ref{fig:Cylinders_interior_gaussian} with the initial disturbance $f(x,y)=e^{-10((x+1.5)^2+y)^2)}$. The error of the GEM solution is $9.7\times 10^{-4}$.}
    \label{fig:Cylinders_exterior_gaussian}
\end{figure}

Figures \ref{fig:SRR_Gaussian}--\ref{fig:SRR_exterior_gaussian} compare the GEM and SEM solutions for a SRR with radius $a=1$\,m and half opening angle $\alpha=\pi/4$ for three initial conditions. These are an interior Gaussian disturbance $f(x,y)=e^{-2(x^2+y^2)}$, an interior dipolar disturbance $f(x,y)=\sin(y)e^{-2(x^2+y^2)}$ and an exterior Gaussian disturbance $f(x,y)=e^{-2((x+2.5)^2+(y+2.5)^2)}$, respectively. The dipolar disturbance was chosen to excite the high Q-factor resonance at $(1.9437 - 0.0294\upi$)\,s$^{-1}$ (i.e.\ the one shown in figure \ref{fig:resonant_modes}(b)). Figures \ref{fig:Cylinders_interior_gaussian}--\ref{fig:Cylinders_exterior_gaussian} compare the GEM and SEM solutions for a square array of four cylinders of radius $a=0.33$\,m with centres $(0.5,0.5)$\,m, $(-0.5,0.5)$\,m, $(0.5,-0.5)$\,m and $(-0.5,-0.5)$\,m, again with three initial conditions. These are an interior Gaussian disturbance $f(x,y)=e^{-10(x^2+y^2)}$, an interior quadrupolar disturbance $f(x,y)=\sin(3x)\sin(3y)e^{-10(x^2+y^2)}$ and an exterior Gaussian disturbance $f(x,y)=e^{-10((x+1.5)^2+y)^2)}$, respectively. The quadrupolar disturbance was chosen to excite the high Q-factor complex resonance at $(8.2132 - 0.0410i)$\,s$^{-1}$ (i.e.\ the one shown in figure \ref{fig:resonant_modes}(d)). In all our results, we set the initial velocity $g(x,y)=0$.

To generate the figures, the truncation frequency in the GEM was taken to be $\omega_{\mathrm{max}}=15$\,s$^{-1}$ for the SRR and $\omega_{\mathrm{max}}=20$\,s$^{-1}$ for the array of cylinders, where larger frequencies were considered due to the more concentrated initial conditions. Likewise, the set of complex frequencies used in the SEM were truncated to $\omega_j^\circ$ values such that $\-1<\Im(\omega_j^\circ)$ and $0<\Re(\omega_j^\circ)<\omega_{\mathrm{max}}$. Inside this region, there are 46 and 11 complex resonant frequencies of the SRR and the array of cylinders, respectively. The spatial discretisation used for the SRR was a square grid with $\Delta x=\Delta y =0.05$\,m and the computational domain was $\mathscr{D}=\{(x,y):|x|,|y|<5\mathrm{m}\}$. The spatial discretisation used for the array of cylinders was a triangulation generated using the Partial Differential Equations Toolbox(TM) in \textsc{MATLAB}(R), with maximum edge length $0.025$\,m. The computational domain was $\mathscr{D}=\{(x,y):|x|,|y|<2\mathrm{m}\}\setminus\Gamma$.

We report the error of the GEM approximation as
\begin{equation}\label{error_eq}
    \mathrm{Error}=\|\mathbf{f}-\boldsymbol{\phi}_0\|_\infty,
\end{equation}
i.e.\ the maximum absolute value of the difference between the GEM solution at $t=0$ and the initial condition at each of the nodes of the spatial discretisation. These values are given in the figure captions. Equation \eqref{error_eq} provides a valid description of the error of the GEM because unlike time-stepping methods, its error does not increase with time. This can be seen from the formula for the GEM \eqref{FD_expansion2}, wherein any error is either due to errors in the spectral amplitude functions $A_m(\omega)$ or in the frequency domain solutions (we assume that this latter source of error is negligible due to the accuracy checks conducted in \textsection\ref{frequency_domain_sec}). Errors in the numerical approximations of $A_m(\omega)$, which are dependent on the spatial and frequency discretisation and the number of incident modes $N_{\mathrm{inc}}$, manifest as errors at $t=0$ where the exact solution (i.e.\ the initial conditions) is known. 

The results show a very good qualitative agreement between the GEM and the SEM after an initial transient, i.e.\ after the resonant modes of the resonator have been excited. This error is due to the approximations in \eqref{td_approx_contour} and \eqref{SEM_formula}, which disregard contributions due to the non-real parts of the integration contour and branch cuts. In evaluations where the initial condition is contained within the resonant cavity (figures \ref{fig:SRR_Gaussian}, \ref{fig:SRR_dipole}, \ref{fig:Cylinders_interior_gaussian} and \ref{fig:Cylinders_quadrupole}), the GEM and SEM solutions quickly begin to agree inside the resonator. As time continues to advance, the qualitative agreement can also be observed in the region surrounding the resonant cavity. In evaluations where the initial condition is exterior to the resonant cavity (figures \ref{fig:SRR_exterior_gaussian} and \ref{fig:Cylinders_exterior_gaussian}), the SEM approximation is initially poor, but becomes qualitatively accurate inside the cavity after the wavefront from the initial condition interacts with the resonator.

\section{Conclusion}\label{conclusion_sec}
This paper has used the GEM and SEM to solve the two-dimensional time-domain wave scattering problem. Two scattering geometries were considered, namely the SRR and the array of sound-hard cylinders, which were solved in \textsection\ref{frequency_domain_sec} using an integral equation/Galerkin method and self-consistent multiple scattering theory, respectively. The GEM, which was introduced in \textsection\ref{GEM_sec}, was approximated using quadrature in \textsection\ref{discrete_GEM_sec}. The resulting discrete GEM was subsequently expressed in the form of matrix multiplication, which is well suited for numerical computation. The SEM was introduced in \textsection\ref{SEM_sec}. There, we showed that the normalisation coefficients of the complex resonant modes, which are in the form of divergent integrals, can be regularised using analytic continuation. Our formula for the normalisation coefficients could be applied to a wide range of resonant-scattering problems. In \textsection\ref{results_sec}, we showed that the agreement between the GEM and SEM was good after an initial transient, i.e.\ after the resonant modes of the resonator have been excited.

The discrete GEM could be readily implemented to solve a wide range of transient wave scattering problems where the frequency-domain solutions are already known. One of the major difficulties in the implementation of the discrete GEM is the selection of the numerical parameters, namely the spatial resolution, the number of frequency quadrature points and the number of incident modes. As we did not know how to choose these optimally, we needed to rely on guesswork and heuristics to reduce the \textit{a posteriori} error to a satisfactory degree. Thus, a better understanding of the convergence properties of the discrete GEM would be incredibly beneficial. Another main difficulty in implementing the discrete GEM is the computational cost of evaluating the frequency-domain solutions. This could be done in parallel using modern high performance computing infrastructure, but we have not explored this here.

\section*{Acknowledgements}
B.W., M.H.M. and F.M. would like to thank the Isaac Newton Institute for Mathematical Sciences and the M{\o}ller Institute at Cambridge (UK) for their support and hospitality during the programme Mathematical theory and applications of multiple wave scattering. This work was supported by EPSRC grant no EP/R014604/1. B.W. and F.M. were supported by a Simons Foundation visiting fellowship to attend the programme.\\
B.W. acknowledges support from a University of Otago Doctoral Scholarship. F.M. acknowledges financial support from Royal Society Te Ap\={a}rangi (Marsden Fund projects 18-UOO-216 and 20-UOO-173) and New Zealand's Antarctic Science Platform (project ANTA1801).

\bibliographystyle{unsrt}
\bibliography{bibfile}

\begin{thebibliography}{10}

\bibitem{martin2006}
Paul Martin.
\newblock {\em Multiple scattering: interaction of time-harmonic waves with N obstacles}.
\newblock Number 107 in Encyclopedia of Mathematics and its Applications. Cambridge University Press, 2006.

\bibitem{wilks2023canonical}
Ben Wilks, Michael~H. Meylan, Fabien Montiel, and Sarah Wakes.
\newblock Generalised eigenfunction expansion and singularity expansion methods for canonical time-domain wave scattering problems, 2023.
\newblock Available at https://arxiv.org/abs/2311.09170.

\bibitem{povzner1953expansion}
Aleksandr~Yakovlevich Povzner.
\newblock On the expansion of arbitrary functions in characteristic functions of the operator -$\delta u+cu$ (in russian).
\newblock {\em Matematicheskii Sbornik}, 74(1):109--156, 1953.

\bibitem{ikebe1960eigenfunction}
Teruo Ikebe.
\newblock Eigenfunction expansions associated with the schroedinger operators and their applications to scattering theory.
\newblock {\em Archive for Rational Mechanics and Analysis}, 5:1--34, 1960.

\bibitem{hazard2007generalized}
Christophe Hazard and Fran{\c{c}}ois Loret.
\newblock Generalized eigenfunction expansions for conservative scattering problems with an application to water waves.
\newblock {\em Proceedings of the Royal Society of Edinburgh Section A: Mathematics}, 137(5):995--1035, 2007.

\bibitem{martin2021time}
Paul~A Martin.
\newblock {\em Time-domain scattering}, volume 180.
\newblock Cambridge University Press, 2021.

\bibitem{meylan_2009}
Michael~H. Meylan.
\newblock Time-dependent linear water-wave scattering in two dimensions by a generalized eigenfunction expansion.
\newblock {\em Journal of Fluid Mechanics}, 632:447–455, 2009.

\bibitem{meylan2009time}
Michael~H Meylan and Rodney~Eatock Taylor.
\newblock Time-dependent water-wave scattering by arrays of cylinders and the approximation of near trapping.
\newblock {\em Journal of Fluid Mechanics}, 631:103--125, 2009.

\bibitem{Meylan2023MWSW03}
Michael~H. Meylan.
\newblock Time-domain calculations for multiple wave scattering.
\newblock presentation given at Isaac Newton Institute Multiple Wave Scattering Workshop 3: Computational methods for multiple scattering, 2023.

\bibitem{baum1971singularity}
Carl~E Baum.
\newblock On the singularity expansion method for the solution of electromagnetic interaction problems.
\newblock {\em Interaction note}, 88(11), 1971.

\bibitem{baum2005singularity}
Carl~E Baum.
\newblock The singularity expansion method.
\newblock In Leopold~B. Felsen, editor, {\em Transient electromagnetic fields}, pages 129--179. Springer, 2005.

\bibitem{pagneux2013trapped}
Vincent Pagneux.
\newblock Trapped modes and edge resonances in acoustics and elasticity.
\newblock In R.V. Craster and J.~Kaplunov, editors, {\em Dynamic Localization Phenomena in Elasticity, Acoustics and Electromagnetism}, pages 181--223. Springer, 2013.

\bibitem{meylan2017extraordinary}
Michael~H Meylan, Mahmood-ul Hassan, and Amna Bashir.
\newblock Extraordinary acoustic transmission, symmetry, blaschke products and resonators.
\newblock {\em Wave Motion}, 74:105--123, 2017.

\bibitem{LINTON200716}
C.M. Linton and P.~McIver.
\newblock Embedded trapped modes in water waves and acoustics.
\newblock {\em Wave Motion}, 45(1):16--29, 2007.
\newblock Special Issue on Localization of Wave Motion.

\bibitem{Kristensen2020}
Philip~Tr{\o}st Kristensen, Kathrin Herrmann, Francesco Intravaia, and Kurt Busch.
\newblock Modeling electromagnetic resonators using quasinormal modes.
\newblock {\em Adv. Opt. Photon.}, 12(3):612--708, Sep 2020.

\bibitem{stoker1957water}
James~Johnston Stoker.
\newblock {\em Water waves: The mathematical theory with applications}.
\newblock New York Interscience Publishers, 1957.

\bibitem{Smith2000}
D.~R. Smith, Willie~J. Padilla, D.~C. Vier, S.~C. Nemat-Nasser, and S.~Schultz.
\newblock Composite medium with simultaneously negative permeability and permittivity.
\newblock {\em Phys. Rev. Lett.}, 84:4184--4187, 5 2000.

\bibitem{llewellyn2010}
Stefan~G Llewellyn~Smith and Anthony~MJ Davis.
\newblock The split ring resonator.
\newblock {\em Proceedings of the Royal Society A: Mathematical, Physical and Engineering Sciences}, 466(2123):3117--3134, 2010.

\bibitem{Hu2011}
Xinhua Hu, C.~T. Chan, Kai~Ming Ho, and Jian Zi.
\newblock {Negative effective gravity in water waves by periodic resonator arrays}.
\newblock {\em Physical Review Letters}, 106(17):174501, 2011.

\bibitem{krynkin2011scattering}
Anton Krynkin, Olga Umnova, Alvin~YB Chong, Shahram Taherzadeh, and Keith Attenborough.
\newblock Scattering by coupled resonating elements in air.
\newblock {\em Journal of Physics D: Applied Physics}, 44(12):125501, 2011.

\bibitem{montiel2017}
F~Montiel, H~Chung, M~Karimi, and N~Kessissoglou.
\newblock An analytical and numerical investigation of acoustic attenuation by a finite sonic crystal.
\newblock {\em Wave Motion}, 70:135--151, 2017.

\bibitem{Bennetts2018}
L.~G. Bennetts, M.~Peter, and R.~Craster.
\newblock {Graded resonator arrays for spatial frequency separation and amplification of water waves}.
\newblock {\em Journal of Fluid Mechanics}, 854:R4, 11 2018.

\bibitem{montiel2020}
Fabien Montiel and Hyuck Chung.
\newblock Planar acoustic scattering by a multi-layered split ring resonator.
\newblock {\em The Journal of the Acoustical Society of America}, 148(6):3698--3708, 2020.

\bibitem{smith2022}
Michael~JA Smith and I~David Abrahams.
\newblock Tailored acoustic metamaterials. part i. thin-and thick-walled helmholtz resonator arrays.
\newblock {\em Proceedings of the Royal Society A}, 478(2262):20220124, 2022.

\bibitem{evans1999trapping}
DV~Evans and R~Porter.
\newblock Trapping and near-trapping by arrays of cylinders in waves.
\newblock {\em Journal of Engineering Mathematics}, 35:149--179, 1999.

\bibitem{maniar1997wave}
H.~D. Maniar and J.~N. Newman.
\newblock Wave diffraction by a long array of cylinders.
\newblock {\em Journal of fluid mechanics}, 339:309--330, 1997.

\bibitem{eatock2007modelling}
Rodney Eatock~Taylor and University of~Oxford.
\newblock On modelling the diffraction of water waves.
\newblock {\em Ship Technology Research}, 54(2):54--80, 2007.

\bibitem{thompson2008new}
Ian Thompson, C.~M. Linton, and R.~Porter.
\newblock A new approximation method for scattering by long finite arrays.
\newblock {\em The Quarterly Journal of Mechanics \& Applied Mathematics}, 61(3):333--352, 2008.

\bibitem{bennetts2022rayleigh}
L.~G. Bennetts and M.~Peter.
\newblock Rayleigh--bloch waves above the cutoff.
\newblock {\em Journal of Fluid Mechanics}, 940:A35, 2022.

\bibitem{EVANS199783}
D.V. Evans and R.~Porter.
\newblock Near-trapping of waves by circular arrays of vertical cylinders.
\newblock {\em Applied Ocean Research}, 19(2):83--99, 1997.

\bibitem{maling2016whispering}
B~Maling and RV~Craster.
\newblock Whispering bloch modes.
\newblock {\em Proceedings of the Royal Society A: Mathematical, Physical and Engineering Sciences}, 472(2191):20160103, 2016.

\bibitem{rossing2015springer}
Thomas~D. Rossing.
\newblock {\em Springer handbook of acoustics}.
\newblock Springer, second edition, 2014.

\bibitem{wilcox1975scattering}
Calvin~H Wilcox.
\newblock {\em Scattering theory for the d'Alembert equation in exterior domains}, volume 442.
\newblock Springer-Verlag, 1975.

\bibitem{hazard2002}
C.~Hazard and M.~Lenoir.
\newblock Surface water waves.
\newblock In Roy Pike and Pierre Sabatier, editors, {\em Scattering}, pages 618--636. Academic Press, 10 2002.

\bibitem{Porter1995a}
R.~Porter and D.~V. Evans.
\newblock {Complementary approximations to wave scattering by vertical barriers}.
\newblock {\em Journal of Fluid Mechanics}, 294:155--180, 1995.

\bibitem{peter_meylan_linton_2006}
Malte~A. Peter, Michael~H. Meylan, and C.~M. Linton.
\newblock Water-wave scattering by a periodic array of arbitrary bodies.
\newblock {\em Journal of Fluid Mechanics}, 548:237–256, 2006.

\bibitem{montiel2015Evolution}
Fabien Montiel, Vernon~A. Squire, and Luke~G. Bennetts.
\newblock Evolution of directional wave spectra through finite regular and randomly perturbed arrays of scatterers.
\newblock {\em SIAM Journal on Applied Mathematics}, 75(2):630--651, 2015.

\bibitem{Amini1995}
S.~Amini and S.M. Kirkup.
\newblock Solution of helmholtz equation in the exterior domain by elementary boundary integral methods.
\newblock {\em Journal of Computational Physics}, 118(2):208--221, 1995.

\bibitem{WANG2004557}
C.D. Wang and M.H. Meylan.
\newblock A higher-order-coupled boundary element and finite element method for the wave forcing of a floating elastic plate.
\newblock {\em Journal of Fluids and Structures}, 19(4):557--572, 2004.

\bibitem{masmoudi1987}
M~Masmoudi.
\newblock Numerical solution for exterior problems.
\newblock {\em Numerische mathematik}, 51(1):87--101, 1987.

\bibitem{kirsch1990}
Andreas Kirsch and Peter Monk.
\newblock Convergence analysis of a coupled finite element and spectral method in acoustic scattering.
\newblock {\em IMA journal of numerical analysis}, 10(3):425--447, 1990.

\bibitem{bateman1954tables}
Harry Bateman.
\newblock {\em Tables of integral transforms}, volume~1.
\newblock McGraw-Hill book company, 1954.

\bibitem{larson2013}
Mats~G Larson and Fredrik Bengzon.
\newblock {\em The finite element method: theory, implementation, and applications}, volume~10.
\newblock Springer Science \& Business Media, 2013.

\bibitem{steinberg1968meromorphic}
Stanly Steinberg.
\newblock Meromorphic families of compact operators.
\newblock {\em Archive for Rational Mechanics and Analysis}, 31:372--379, 1968.

\bibitem{WOLGAMOT2017232}
H.~A. Wolgamot, M.~H. Meylan, and C.~D. Reid.
\newblock Multiply heaving bodies in the time-domain: Symmetry and complex resonances.
\newblock {\em Journal of Fluids and Structures}, 69:232--251, 2017.

\bibitem{kowalczyk2015complex}
Piotr Kowalczyk.
\newblock Complex root finding algorithm based on delaunay triangulation.
\newblock {\em ACM Transactions on Mathematical Software (TOMS)}, 41(3):1--13, 2015.

\bibitem{Kowalczyk2018}
Piotr Kowalczyk.
\newblock Global complex roots and poles finding algorithm based on phase analysis for propagation and radiation problems.
\newblock {\em IEEE Transactions on Antennas and Propagation}, 66(12):7198--7205, 2018.

\bibitem{delves1967numerical}
LM~Delves and JN~Lyness.
\newblock A numerical method for locating the zeros of an analytic function.
\newblock {\em Mathematics of computation}, 21(100):543--560, 1967.

\bibitem{meylan2002parallel}
Michael~H Meylan and Lutz Gross.
\newblock A parallel algorithm to find the zeros of a complex analytic function.
\newblock {\em ANZIAM Journal}, 44:E236--E254, 2002.

\bibitem{abramowitz1988handbook}
Milton Abramowitz, Irene~A Stegun, and Robert~H Romer.
\newblock Handbook of mathematical functions with formulas, graphs, and mathematical tables, 1988.

\bibitem{watson1922treatise}
George~Neville Watson.
\newblock {\em A treatise on the theory of Bessel functions}.
\newblock Cambridge University Press, 2 edition, 1944.

\end{thebibliography}

\end{document}